\documentclass[a4paper,11pt]{article}
\usepackage{graphicx}
\usepackage{amsfonts}
\usepackage{amsmath}
\usepackage[utf8]{inputenc}
\usepackage{url}
\usepackage{color}

\newcommand{\Pd}{\mbox{$\mathcal{P}_2(\mathbb{R}^d)$} }
\newcommand{\Wd}{\mbox{$\mathcal{W}_2$} }
\newcommand{\Wdd}{\mbox{$\mathcal{W}_2^2$}}
\newcommand{\Read}{\mbox{$ \mathbb{R}^d$} }

\newcommand{\Rea}{\mbox{$\mathbb{R}$}}

\newcommand{\Prob}{\mbox{$\mathbb{P}$}}

\newcommand{\conv}{\rightarrow}

\newcommand{\convw}{\rightarrow_w}

\newtheorem {Proposition}{Proposition} [section]

\newtheorem {Theorem}[Proposition]{Theorem}
\newtheorem {Corollary}[Proposition]{Corollary}
\newtheorem {Remark}[Proposition]{Remark}
\newtheorem {Example}{Example}[section]
\newtheorem {Definition}{Definition}[section]

\begin{document}
\title{\sc A fixed-point approach to barycenters in Wasserstein space\footnote{Research partially supported by the
Spanish Ministerio de Econom\'{\i}a y Competitividad, grants  
MTM2014-56235-C2-1-P and MTM2014-56235-C2-2, and  by Consejer\'{\i}a de Educaci\'on de la Junta de Castilla y Le\'on, grant VA212U13.}}

\author{Pedro C. \'Alvarez-Esteban$^{1}$, E. del Barrio$^{1}$, J.A. Cuesta-Albertos$^{2}$\\ and C. Matr\'an$^{1}$ \\
$^{1}$\textit{Departamento de Estad\'{\i}stica e Investigaci\'on Operativa and IMUVA,}\\
\textit{Universidad de Valladolid} \\ $^{2}$ \textit{Departamento de
Matem\'{a}ticas, Estad\'{\i}stica y Computaci\'{o}n,}\\
\textit{Universidad de Cantabria}}
\maketitle

\begin{abstract}
Let $\mathcal{P}_{2,ac}$ be the set of Borel probabilities on $\mathbb{R}^d$ with finite second moment and absolutely continuous 
with respect to Lebesgue measure.
We consider the problem of finding the barycenter (or Fr\'echet mean) of a finite set of probabilities $\nu_1,\ldots,\nu_k \in  \mathcal{P}_{2,ac}$ 
with respect to the $L_2-$Wasserstein metric. For this task we introduce an operator on $\mathcal{P}_{2,ac}$ related to the optimal transport 
maps pushing forward any $\mu \in  \mathcal{P}_{2,ac}$ to $\nu_1,\ldots,\nu_k$. Under very general conditions we prove that the barycenter 
must be a fixed point for this operator and introduce an iterative procedure which consistently approximates
the barycenter. The procedure allows effective computation of barycenters  in any location-scatter
family, including the Gaussian case. In such cases the barycenter must belong to the family, thus it is characterized 
by its mean and covariance matrix. While its mean is just the weighted mean of the means of the probabilities, 
the covariance matrix is characterized in terms  of their covariance matrices  $\Sigma_1,\dots,\Sigma_k$ through 
a nonlinear matrix equation.  The performance of the iterative procedure in this case is illustrated through
numerical simulations, which show fast convergence towards the barycenter.
\end{abstract}


\noindent
{\it A.M.S. classification:} {\sc Primary } 60B05. {\sc Secondary: } 47H10, 47J25, 65D99.

\section{Introduction.}
Let us consider a set $\{x_1,\dots,x_k\}$ of elements in a  certain space and associated weights, $\lambda_1,\dots,\lambda_k$, 
satisfying $\lambda_i> 0$, $\sum_{i=1}^n \lambda_i=1$,  interpretable as a quantification of  the relative importance or presence of these elements.
The suitable choice of an element in the space to represent  that set  is an old problem present in 
many different settings. The weighted mean being the best known choice, it enjoys many nice properties that allow us to
consider it a very good representation of  elements in an Euclidean space. Yet, it can be highly undesirable 
for representing shaped objects like functions or matrices with some particular structure. The Fr\'echet mean or barycenter is a 
natural extension arising from the consideration of minimum dispersion character of the mean, when the space has a 
metric structure  which, in some cases, may overcome these difficulties. Like the mean, if $d$ is a distance in the reference space $E$, a barycenter $\bar x$ is determined by the relation
$$\sum_{i=1}^k\lambda_i d^2(x_i,\bar x)=\min\left\{\sum_{i=1}^k \lambda_i d^2(x_i,y), \ y \in E\right\}.$$
In the last years Wasserstein spaces have focused the interest of researchers from very different 
fields (see, e.g., the monographs \cite{AGS}, \cite{Villani} or \cite{Villani2}), leading in 
particular to the natural consideration of Wasserstein barycenters beginning with \cite{AguehCarlier}. 
This appealing concept shows a high potential for application, already considered in Artificial Intelligence 
or in Statistics (see, e.g., \cite{Cuturi}, \cite{Benamou}, \cite{Carlier}, \cite{Le Gouic}, \cite{Bishop} 
or \cite{wideconsensus}). The main drawback being the difficulties for its effective computation, 
several of these papers (\cite{Cuturi}, \cite{Benamou}, \cite{Carlier}) are mainly devoted to this 
hard goal. In fact, the ($L_2-$)Wasserstein distance between probabilities, which is the framework 
of this paper, is easily characterized and computed for probabilities on the real line, but 
there is not a similar, simple closed expression for its computation in higher dimension. 
A notable exception arises from Gelbrich's bound and some extensions (see \cite{Gelbrich}, 
\cite{{Cuesta96}}, \cite{Cuesta-Albertos1993}) that allow the computation of the distances between 
normal distributions or between probabilities in some parametric (location-scatter) families. 
For multivariate Gaussian distributions  in particular, in  \cite{AguehCarlier}, the barycenter has been characterized in 
terms of a fixed point  equation involving the covariance matrices in a nonlinear way 
(see     (\ref{matrixeq}) below) but, to the best of our knowledge, feasible consistent algorithms to 
approximate the solution have not been proposed yet.

The approach in \cite{AguehCarlier} for the characterization of the barycenter in the Gaussian setting resorted 
to duality arguments and to Brouwer's fixed-point theorem. Here we take a different approach which is not constrained
to the Gaussian setup. We introduce an operator associated to the transformation of a probability measure 
through weighted averages of optimal transportations to the set $\{\nu_i\}_{i=1}^k$ of target distributions. 
This operator is the real core of our approach. We show (see Theorem \ref{p1} and Proposition \ref{basicineq} below) 
that, in very general situations, barycenters must be fixed points of the above mentioned operator. We also show (Theorem \ref{PrevioConsistencia}) 
how this operator can be used to define a consistent iterative scheme for the approximate computation of barycenters. 
Of course, the practical usefulness of the iteration will depend on the difficulties arising from the computation of the optimal transportation maps
involved in the iteration. The case of Gaussian probabilities is a particularly convenient setup for our iteration. 
We provide a self-contained approach to  barycenters in this Gaussian framework based on first principles of 
optimal transportation and some elementary matrix analysis. This leads also to the  characterization (\ref{matrixeq}) and,
furthermore, it yields sharp bounds on the covariance matrix of the barycenter which are of independent interest. 
We prove (Theorem \ref{normaliteration}) that our iteration provides a consistent approximation to barycenters in this
Gaussian setup.
We also notice that all the results given for the Gaussian 
family are automatically extended to location-scatter families  (see Definition \ref{loc.scale} below). 
Finally, we illustrate the performance of the iteration through numerical simulations. These show fast convergence towards the barycenter,
even in problems involving a large number of distributions or high-dimensional spaces. 

The remaining sections of this paper are organized as follows. Section 2 gives a succint account of some basic facts about optimal transportation and
Wasserstein metrics and introduces the barycenter problem with respect to these metrics. The section contains pointers to the most relevant references
on the topic. Section 3 contains the core of the paper, introducing the operator $G$ in (\ref{Gdefinition}), showing the connection
between barycenters and fixed points of $G$ and presenting the iterative scheme for approximate computation of
barycenters. The Gaussian and location-scatter cases are analyzed in Section 4, while Section 5 presents some numerical simulations.
We conclude this Introduction with some words on notation. Throughout the paper our space 
of reference is the Euclidean space $\Read$. With $\|x\|$ we 
denote the usual norm and with $x\cdot y$ the inner product. For a matrix $A$, $A^t$ will 
denote the corresponding transpose matrix, $\det(A)$ the determinant and $\mbox{Tr}(A)$ 
the trace. $Id$ will be indistinctly used as the identity  map and as the $d\times d$ identity 
matrix, while $ {\cal M}_{d\times d}^+$ will denote the set of $d\times d$ (symmetric) positive definite matrices The space where we consider the Wasserstein distance is $\mathcal{P}_2\equiv 
\mathcal{P}_2(\Read)$, the set of Borel probabilities on $\mathbb{R}^d$ with finite 
second moment. The related set  $\mathcal{P}_{2,ac}\equiv \mathcal{P}_{2,ac}(\Read)$ will denote 
the subset of $\mathcal{P}_2(\Read)$ containing the probabilities that are absolutely continuous 
with respect to Lebesgue measure. The probability law of a random vector $X$ will be represented 
by  $\mathcal L(X)$.

\section{Wasserstein spaces and barycenters.}

Given $\mu,\nu \in  \mathcal{P}_2(\Read)$, the ($L_2-$)Wasserstein distance 
between them is defined as
\begin{equation}\label{distancia}
\Wd(\mu,\nu):=\left(\inf\left\{\int\|x-y\|^2d\pi(x,y), \pi \in \mathcal{P}_2
(\Read \times \Read) \mbox{ with marginals } \mu,\nu \right\}\right)^{1/2}.
\end{equation}
It is well known that this distance metrizes weak convergence of probabilities plus convergence of second moments, that is,
\begin{equation}\label{caracter}
\Wd(\mu_n,\mu)\conv 0 \iff \mu_n \convw \mu \ \mbox{ and } \int\|x\|^2d\mu_n(x)\conv \int\|x\|^2d\mu(x).
\end{equation}
We refer to \cite{Villani} for details. In the one-dimensional case $\Wd(\mu,\nu)$ is simply
the $L_2$ distance between the quantile functions of $\mu$ and $\nu$, allowing computation. In 
the multivariate case there exist just a few general results that 
can be used to simplify problems. An important example, which allows to deal with certain problems 
by considering only the case of centered probabilities is the following,
\begin{equation}\label{centering}
\Wdd(P,Q)=\|m_P-m_Q\|^2+\Wdd(P^*,Q^*),
\end{equation}
where $P$ and $Q$ are arbitrary probabilities in $\mathcal{P}_{2}(\Read)$, with 
respective means $m_P,m_Q$, and $P^*,Q^*$ are the corresponding centered in mean probabilities.

In any case, it is well known that the infimum in (\ref{distancia}) is attained. A distribution $\pi$ with marginals $\mu$ and $\nu$ attaining that infimum is called an optimal coupling  of $\mu$ and $\nu$. Moreover, if $\mu$  vanishes 
on sets of dimension $d-1$, in particular if $\mu \in \mathcal{P}_{2,ac}(\Read)$, then 
there exists an optimal transport map, $T$, transporting (pushing forward) $\mu$ to $\nu$.
This important result is the final product of  partial results and successive improvements obtained in several papers including Knott and Smith \cite{Knott1}, Brenier \cite{Brenier1}, 
\cite{Brenier2},  Cuesta-Albertos and 
Matr\'an \cite{Cuesta}, R\"uschendorf and Rachev  \cite{Ruschen} and McCann \cite{McCann}. We include 
for the ease of reference a convenient version  (see Theorem 2.12 in Villani's book \cite{Villani}) 
and also refer to \cite{Villani2} for the general theory on  Wasserstein distances 
and the transport problem. We recall that for a lower semicontinuous convex function 
$\varphi$, the conjugate function is defined as $\varphi^*(y):= \sup \{x\cdot y-\varphi(x),  \ x\in \Read\}$.

\begin{Theorem}\label{BCMRR}
Let $\mu,\nu \in  \mathcal{P}_2(\Read)$ and let $\pi=\mathcal L(X,Y)$ be the joint distribution of a pair $(X,Y)$ of $\Read-$valued random vectors with probability laws $\mathcal L(X)=\mu$ and $\mathcal L(Y)=\nu$. 

\noindent
a) The probability distribution $\pi$ is an optimal coupling of $\mu$ and $\nu$ if and only if  there exists a convex lower semi-continuous function $\varphi$ such that $Y\in \partial\varphi(X)$  a.s.

\noindent
b) If we assume that $\mu$ does not give mass to sets of 
dimension at most $d-1$, then there is a unique  optimal coupling $\pi$
of $\mu$ and $\nu$, that can be characterized as the unique solution to the Monge transportation 
problem --an optimal transport map-- $T$, i.e.: $\pi=\mu\circ (Id,T)^{-1}$ (or $Y=T(X)$ a.s.),  and
$$
\Wdd(\mu,\nu)=\int \|x-T(x)\|^2d\mu(x)= 
\inf \left\{\int\|x-S(x)\|^2d\mu(x), \mbox{ where $S$ satisfies  } \nu=\mu\circ S^{-1} \right\}.
$$  

Such a map is characterized as $T=\nabla \varphi$ $\mu$- a.s., the $\mu$-a.s. unique  function that maps $\mu$ to $\nu$ and that is the gradient of a lower semicontinuous, convex function $\varphi$.
Moreover, if $\nu$ also does not give mass to sets of dimension at most $d-1$, then for $\mu$-a.e. $x$ and $\nu$-a.e. $y$, 
$$\nabla\varphi^*\circ \nabla \varphi(x)=x, \ \  \nabla\varphi \circ \nabla \varphi^*(y)=y,$$
and $\nabla \varphi^*$ is the ($\nu$-a.s.) unique optimal transport map from $\nu$ to $\mu$
and the unique  function that maps $\nu$ to $\mu$ and that is the  gradient of a convex, lower semicontinuous function. 

\end{Theorem}

\begin{Remark}\label{cyclicmon}{\em
{Concerning optimal couplings, from a) it easily follows that if $\pi_j=\mathcal L(X,Y_j), j=1,\dots,k,$ are optimal couplings of $\mu$ and $\nu_j$, for positive weights $\lambda_j$ satisfying $\sum_{j=1}^k\lambda_j=1$, the probability $\pi=\mathcal L (X,\sum_{j=1}^k\lambda_jY_j)$ is an optimal coupling for $\mu$ and $\mathcal L (\sum_{j=1}^k\lambda_jY_j)$.}
Another remarkable consequence of the theorem is that optimality of a map is a 
characteristic that does not depend on the transported measures. In 
other words, if $T$ is an optimal map for transporting $\mu$ to $\nu$ and $\mu^*,\nu^* 
\in  \mathcal{P}_2(\Read)$ are such that $\nu^*=\mu^*\circ T^{-1}$ and the support of $\mu^*$ is 
contained in that of $\mu$, then $T$ is also an optimal transport map from $\mu^*$ to 
$\nu^*$: $\Wdd(\mu^*,\nu^*)=\int \|x-T(x)\|^2d\mu^*(x).$ This fact allows the computation 
of the distance between some probabilities when we know that they can be related through optimal 
maps. It must be also noted that composition 
of optimal maps does not generally preserve optimality, but positive linear combinations and 
point-wise limits of optimal maps keep optimality. 
}
\end{Remark}

Special mention among the class of optimal maps must be given to the class of positive definite affine 
transformations. This is done in the following theorem,
a version of  Theorem 2.1 in Cuesta-Albertos et al. \cite{Cuesta96}, that suffices 
for our purposes here, giving an additional perspective to Gelbrich's bound (\ref{cotaGelbrich}) 
(see \cite{Gelbrich}). 

\begin{Theorem}\label{Gelbrich2}
Let $P$ and $Q$ be probabilities in \Pd with means $m_P, m_Q$ and  covariance matrices $\Sigma_P, \Sigma_Q$. If $\Sigma_P$ is assumed nonsingular, then 
\begin{equation}\label{cotaGelbrich}
\Wdd(P,Q)\geq \|m_P-m_Q\|^2+ \mbox{\em Tr}\left(\Sigma_P+\Sigma_Q-2\left(\Sigma_P^{1/2}\Sigma_Q\Sigma_P^{1/2}\right)^{1/2}\right).
\end{equation}
Moreover, equality holds if and only if the map $T(x)=(m_Q-m_P)+Ax$ transports $P$ to $Q$, where $A$ is the positive semidefinite
matrix given by
\[
A=\Sigma_P^{-1/2}\left(\Sigma_P^{1/2}\Sigma_Q\Sigma_P^{1/2}\right)^{1/2}\Sigma_P^{-1/2}.
\]
\end{Theorem}

This theorem allows an easy generalization of the results involving optimal transportation 
results between Gaussian probabilities to a wider setting of probability families that we introduce now. 
Recall that $ {\cal M}_{d\times d}^+$ is the set of $d\times d$ (symmetric) positive definite matrices.

\begin{Definition}\label{loc.scale}
Let $\bf{X}_0$ be a random vector with probability law  ${\bf P}_0 \in \mathcal{P}_{2,ac}(\Read)$. 
The family $\mathcal{F}({\bf P}_0):=\{\mathcal{L}(A{\bf X}_0+m), A\in {\cal M}_{d\times d}^+, m\in\Read\}$  
of probability laws induced by  positive definite affine transformations from ${\bf P}_0$ will be 
called a location-scatter family. 
\end{Definition}

Of course a location-scatter family  $\mathcal{F}({\bf P}_0)$ can be parameterized through the 
parameters $m$ and $A$ that appear in the definition. Note, however, that, if $m_0$ and $\Sigma_0$ 
are the mean and covariance matrix of ${\bf P}_0$, the family can be also defined from  
${\bf P}^*_0:=\mathcal{L}(\Sigma_0^{-1/2}({\bf X}_0-m_0))$, with mean $0$ and covariance 
matrix $Id$. This allows to parameterize the family through the mean and the covariance 
matrix of the laws in the family, a fact that will be assumed throughout without additional 
mention. With this assumption, a probability law in the 
family $\mathcal{F}({\bf P}_0)$ will be denoted in terms of its mean, $m$, and its covariance matrix, $\Sigma$, by $\mathbb{P}_{m,\Sigma}$, and, as a consequence of Theorem \ref{Gelbrich2},  we have 
\begin{eqnarray*} \nonumber
\Wdd(\mathbb{P}_{m_1,\Sigma_1},\mathbb{P}_{m_2,\Sigma_2})&=& \Wdd(N(m_1,\Sigma_1),N(m_2,\Sigma_2))
\\ 
&=&\|m_1-m_2\|^2+ \mbox{Tr}\left(\Sigma_1+\Sigma_2-2\left(\Sigma_1^{1/2}\Sigma_2\Sigma_1^{1/2}\right)^{1/2}\right).
\end{eqnarray*}

The central problem of this work involves  probabilities  $\nu_1,\ldots,\nu_k \in \mathcal{P}_2(\Read)$ 
and fixed weights $\lambda_1,\ldots,\lambda_k$ that are positive
real numbers such that $\sum_{i=1}^k\lambda_i=1$. For an arbitrary $\mu\in \mathcal{P}_2(\Read)$, 
we will consider the functional 
$$
V(\mu):=\sum_{i=1}^k\lambda_i\Wdd(\mu,\nu_i).
$$

\begin{Definition}\label{baricentro}
If $\bar \mu \in \mathcal{P}_2(\mathbb{R}^d)$ is such that $$V(\bar{\mu})=\min_{\mu\in\mathcal{P}_2}V(\mu),$$
then we say that  $\bar \mu$ is a \textit{barycenter} (with respect to Wasserstein distance)
of $\nu_1,\ldots,\nu_k$. 
\end{Definition}
Through the paper we will maintain the notation $\bar \mu$ for the barycenter. Existence and 
uniqueness of barycenters have been considered
in \cite{AguehCarlier}. For the sake of completeness, we give here a succinct argument to
prove these existence and uniqueness. Note first that given probabilities $\mu, \nu_1,\ldots,\nu_k$ on $\mathbb{R}^d$ 
and joint probabilities $\pi_j$ on $\mathbb{R}^d\times \mathbb{R}^d$  with marginals $\mu$ and $\nu_j$ we can always find
a joint probability $\tilde{\pi}$ on $(\mathbb{R}^d)^{k+1}$ such that the marginal of $\tilde{\pi}$
over factors $1$ and $j+1$ equals $\pi_j$, $j=1,\ldots,k$ (this fact is often invoked as ``gluing lemma", see e.g. \cite{Villani2}). Hence, we have that
$$\inf_\mu V(\mu)=\inf_{\tilde{\pi}\in\Pi(\cdot,\nu_1,\ldots,\nu_k)} \int_{(\mathbb{R}^d)^{k+1}}
\sum_{j=1}^k\lambda_j \|x-x_j\|^2 d\tilde{\pi}(x,x_1,\ldots,x_k),$$
where $\Pi(\cdot,\nu_1,\ldots,\nu_k)$ denotes the set of probabilities on $(\mathbb{R}^d)^{k+1}$
whose last $k$ marginals are $\nu_1,\ldots,\nu_k$. Of course, for fixed $x_1,\ldots,x_k$
we have $\sum_{j=1}^k\lambda_j \|x-x_j\|^2\geq \sum_{j=1}^k\lambda_j \|\bar{x}-x_j\|^2$ with
$\bar{x}=\sum_{j=1}^k\lambda_j x_j$. This means that among all $\tilde{\pi}$ with the same
(joint) marginal distribution over the last $k$ factors the functional 
$\int_{(\mathbb{R}^d)^{k+1}}
\sum_{j=1}^k\lambda_j \|x-x_j\|^2 d\tilde{\pi}(x,x_1,\ldots,x_k)$ is minimized when
$\tilde{\pi}$ is concentrated on the set $\big\{(x,x_1,\ldots,x_k): x=\sum_{j=1}^k \lambda_j x_j\big\}$. This means that
\begin{equation}\label{problema}
\inf_\mu V(\mu)=\inf_{\pi\in\Pi(\nu_1,\ldots,\nu_k)} \int_{(\mathbb{R}^d)^{k}}
\sum_{j=1}^k\lambda_j \|\bar{x}-x_j\|^2 d{\pi}(x_1,\ldots,x_k),
\end{equation}
with $\Pi(\nu_1,\ldots,\nu_k)$ denoting the set of probabilities on $(\mathbb{R}^d)^{k}$
with marginals equal to $\nu_1,\ldots,\nu_k$, and a barycenter is the law induced by the map
$(x_1,\ldots,x_k)\mapsto \bar{x}=\sum_{j=1}^k \lambda_j x_j$ from $\bar{\pi}$, a minimizer for the right-hand side in
(\ref{problema}), provided it exists. Now, write $\bar{V}$ for the minimal value in 
(\ref{problema}) and assume that $\{\pi_n\}_n \subset \Pi(\nu_1,\ldots,\nu_k)$ is a minimizing sequence, that is, that
$$\int_{(\mathbb{R}^d)^{k}}
\sum_{j=1}^k\lambda_j \|\bar{x}-x_j\|^2 d{\pi_n}(x_1,\ldots,x_k)\to \bar{V},
$$
as $n\to\infty$. The fact that the marginals of $\pi_n$ are fixed implies that $\{\pi_n\}_n$ is a tight sequence.
Hence, by taking subsequences if necessary, we can assume that $\{\pi_n\}_n$ converges weakly, say to $\bar{\pi}\in
\Pi(\nu_1,\ldots,\nu_k)$. A standard uniform integrability argument (see, e.g., the proof of Theorem \ref{p1} below) 
shows that $$\int_{(\mathbb{R}^d)^{k}}
\sum_{j=1}^k\lambda_j \|\bar{x}-x_j\|^2 d{\pi_n}(x_1,\ldots,x_k)\to \int_{(\mathbb{R}^d)^{k}}
\sum_{j=1}^k\lambda_j \|\bar{x}-x_j\|^2 d{\bar{\pi}}(x_1,\ldots,x_k)=\bar{V},$$
hence, a minimizer exists for the multimarginal problem and, as a consequence, $\bar{\mu}=\bar{\pi}\circ T^{-1}$
with $T(x_1,\ldots,x_k)=\sum_{j=1}^k \lambda_j x_j$, is a barycenter, that is, a minimizer of $V$. Strict convexity of the
map $\mu\mapsto \mathcal{W}_2^2(\mu,\nu)$ when $\nu$ has a density (see Corollary 2.10 in \cite{AEetal})
implies that $V$ is strictly convex if at least one of $\nu_1,\ldots,\nu_k$ has a density and yields uniqueness of
the barycenter. As we said, this existence and uniqueness results for the barycenter are contained in \cite{AguehCarlier} but 
this approach is, arguably, more elementary and shorter.

\medskip
Beyond the existence and uniqueness results and apart from the case of probability distributions 
on the real line, the computation of barycenters is a hard problem. The following theorem 
(Theorem 6.1 in \cite{AguehCarlier}) could be an 
exception because it characterizes the barycenter for nonsingular multivariate Gaussian distributions  and, then, 
for distributions in the same location-scatter family.

\begin{Theorem}\label{casonormal}
Let $\nu_1,\ldots,\nu_k$ be Gaussian distributions with respective means $m_1,\ldots,m_k$ and nonsingular
covariance matrices $\Sigma_1,\ldots,\Sigma_k$. The barycenter of $\nu_1,\ldots,\nu_k$ with
weights $\lambda_1,\ldots,\lambda_k$ is the Gaussian distribution with mean $\bar m=\sum_{i=1}^k
\lambda_i m_i$ and covariance matrix $\bar \Sigma$ defined as the only positive definite matrix 
satisfying the equation
\begin{equation}\label{matrixeq}
S=\sum_{i=1}^k\lambda_i\left(S^{1/2}\Sigma_iS^{1/2}\right)^{1/2}.
\end{equation}
\end{Theorem}

Parts of this result were already outlined in Knott and Smith \cite{Knott} and further developed  
in R\"uschendorf and Uckelmann \cite{Ruschen2}. 
However, by itself it is far from allowing an effective computational method for the 
barycenter.
In the next section we introduce an iterative procedure that, under some general assumptions, produces
convergent approximations to $\bar{\mu}$.

\section{A fixed-point approach to Wasserstein barycenters.}

We introduce in this section a map $G: \mathcal{P}_{2,ac}(\Read)\to \mathcal{P}_{2,ac}(\Read)$ whose 
fixed points are, under mild assumptions, the barycenters of $\nu_1,\ldots,\nu_k\in\mathcal{P}_{2,ac}(\Read)$
with weights $\lambda_1,\ldots\lambda_k$ in the sense of Definition \ref{baricentro}. With this goal, 
consider now an additional $\mu\in\mathcal{P}_{2,ac}(\Read)$. From Theorem \ref{BCMRR}  there exist 
($\mu$-a.s. unique) optimal transportation maps $T_1,\ldots,T_k$ such that
$\mathcal{L}(T_j(X))=\nu_j$, where $X$ is a random vector such that $\mathcal{L}(X)=\mu$, and 
$\mathcal{W}_2^2(\mu,\nu_j)=E\|X-T_j(X) \|^2$, $j=1,\ldots,k$.
Then, we define
\begin{equation}\label{Gdefinition}
G(\mu):=\mathcal{L}\Big(\sum_{j=1}^k \lambda_j T_j(X)\Big).
\end{equation}
We will show the connection between barycenters and fixed points of $G$ and also how we can use the 
transform $G$ to define a consistent, iterative procedure for the approximate 
computation of $\bar{\mu}$. First, we prove some basic properties of $G$.

\begin{Theorem}\label{p1} If $\nu_j$ has a density, $j=1,\ldots,k$, then  
$G$ maps $\mathcal{P}_{2,ac}(\Read)$ into $\mathcal{P}_{2,ac}(\Read)$ and it is continuous for the $\mathcal{W}_2$ metric. 
\end{Theorem}

\medskip
\noindent \textbf{Proof.} From Theorem \ref{BCMRR} we know that $T_j=\nabla \varphi_j$ where $\varphi_j$ is a convex,
lower semicontinuous function. Moreover, since $\nu_j\in\mathcal{P}_{2,ac}(\Read)$, denoting by $\varphi_j^*(y):= \sup \{x\cdot y-\varphi_j(x)\}$ the conjugate function, it holds 
$\nabla \varphi_j^* \circ\nabla\varphi_j(x)=x$ $\mu$-a.s. (in particular, $\varphi_j(x)$ is injective in a set of total $\mu$-measure).
By Alexandrov's Theorem (see, e.g., Theorem 3.11.2 in \cite{Niculescu} or Theorem 1 in Section 6.4 in \cite{Evans}) we have that for $\mu$-almost every $x$ there is a symmetric, 
positive semidefinite matrix, which we denote $\nabla^2 \varphi_j(x)$ such that for every
$\varepsilon>0$ there exists $\delta>0$ such that $\|y-x\|\leq \delta$ implies
$$\sup_{z\in\partial \varphi_j(x)}\|z-\nabla \varphi_j(x)-\nabla^2 \varphi_j(x) (y-x)\|\leq \varepsilon \|y-x\|.$$ 
But then, the fact that $\nu_j$ has a density and Lemma 5.5.3 in \cite{AGS} imply that $\nabla^2 \varphi_j(x)$ is $\mu-$a.s. nonsingular, hence
positive definite. Thus, writing $\varphi=\sum_{j=1}^k \lambda_j \varphi_j$ we have that
$\nabla \varphi(x)=\sum_{j=1}^k \lambda_j \nabla \varphi_j(x)$ and for $\mu$-a.e. $x$ we have that 
that for every
$\varepsilon>0$ there exists $\delta>0$ such that $\|y-x\|\leq \delta$ implies
\begin{equation}\label{taylor2}
\sup_{z\in\partial \varphi(x)}\|z-\nabla \varphi(x)-\nabla^2 \varphi(x) (y-x)\|\leq \varepsilon \|y-x\|,
\end{equation}
where $\nabla^2 \varphi(x)=\sum_{j=1}^k \lambda_j \nabla^2\varphi_j(x)$ is symmetric,
positive definite, hence, nonsingular. We claim that $\nabla \varphi$ is injective outside a $\mu$-negligible set. To see this, choose $x$ satisfying (\ref{taylor2}) and $y\ne x$ such that $\varphi$ is differentiable at
$y$ and consider the convex function $F(s)=\varphi((1-s)x+sy)$. Observe that $F'(0)=\nabla \varphi(x) \cdot (y-x)$
and that for $z\in\partial \varphi((1-s)x+sy)$ we have $z\cdot (y-x)\leq F'(s+)\leq F'(1)=\nabla \varphi(y) \cdot (y-x)$
(this follows, for instance, from Lemma 3.7.2, p 129 in \cite{Niculescu}). Now from
positive definiteness of $\nabla^2 \varphi(x)$ we know that, for some $\rho>0$, $v^t\nabla^2 \varphi(x) v\geq \rho \|v\|^2$
for every $v\in\mathbb{R}^d$. Take $\varepsilon\in (0,\rho/2)$ and $s>0$ small enough to ensure, using (\ref{taylor2}),
that for $z\in\partial \varphi((1-s)x+sy)$, $\| z-\nabla \varphi(x)-s \nabla^2 \varphi(x) (y-x)\|\leq s\varepsilon \|y-x\|$. But then
$$| s(y-x)\cdot (z-\nabla\varphi(x))-s^2 (y-x)^t \nabla^2 \varphi(x) (y-x)|\leq s^2 \varepsilon \|y-x\|^2,$$
which entails 
$$s(y-x)\cdot (z-\nabla\varphi(x))\geq s^2 \|y-x\|^2 \Big( \frac{(y-x)^t \nabla^2 \varphi(x) (y-x)} {\|y-x\|^2}-\varepsilon\Big)
\geq \frac  \rho 2 s^2 \|y-x\|^2>0.$$
As a consequence, $F'(0)=\nabla \varphi(x) \cdot (y-x)< z \cdot (y-x)\leq F'(1)=\nabla \varphi(y) \cdot (y-x)$.
This implies that $\nabla \varphi(x)\ne \nabla\varphi(y)$, that is, $\nabla \varphi$ is injective in a set of total $\mu$-measure.
We can apply now Lemma 5.5.3 in \cite{AGS} with the fact that  $\nabla^2 \varphi(x)$ is $\mu$-a.s. nonsingular to conclude
that $G(\mu)$ has a density. The fact that $G(\mu)=\mathcal{L}(\sum_{j=1}^k \lambda_j X_j)$ with $X_j=T_j(X)$
having law $\mu_j$ implies that $G(\mu)$ has finite second moment, which completes the proof of the first claim.

\medskip
Turning to continuity, assume that $\mu,\{\mu_n \}_n\in\mathcal{P}_{2,ac}(\Read)$ 
satisfy $\mathcal{W}_2(\mu_n,\mu)\to 0$. Write $T_{j}$ (resp. $T_{n,j}$) for the optimal
transportation map from $\mu$ (resp. $\mu_n$) to $\nu_j$, $j=1,\ldots,k$. According to Skorohod's representation theorem (see e.g. Theorem 11.7.2 in \cite{Dudley}), we can consider random vectors $X, X_n, n=1,...$ such that $X_n \to X$ a.s., where $X$ has law $\mu$ and  $X_n$ having law
$\mu_n$. Then, by Theorem 3.4 in \cite{Cuesta2}, also $T_{n,j}(X_n)\to T_j(X)$ a.s. for $j=1,\dots,k$, hence $\sum_{j=1}^k \lambda_j T_{n,j}(X_n) \to \sum_{j=1}^k \lambda_j T_{j}(X)$ a.s., that implies the convergence $\mathcal{L}(\sum_{j=1}^k \lambda_j T_{n,j}(X_n)) \convw \mathcal{L}(\sum_{j=1}^k \lambda_j T_{j}(X)).$

Now, the fact that the families $\{\|T_{n,j}(X_n)\|^2\}_n$ are uniformly integrable (indeed the law of $T_{n,j}(X_n)$ is $\nu_j$, fixed)
entails the uniform integrability of  $\left\{ \left\| \sum_{j=1}^k \lambda_j T_{n,j}(X_n)\right\|^2\right\}_n$ and shows that $\mathcal{W}_2(G(\mu_n),G(\mu))\to 0$ thus, by characterization (\ref{caracter}),
finishing the proof.
\quad $\Box$

\begin{Remark}\label{boundeddensity}{\em
Let us additionally assume the following:
\begin{equation}\label{asum1}
\nu_1,\ldots,\nu_k\in\mathcal{P}_{2,ac}(\Read) \mbox{ and  at least one of them has a 
bounded density}
\end{equation}
Under these assumptions there is a unique barycenter which has a (bounded) density,
see Theorem 5.1 in \cite{AguehCarlier}, and the first conclusion in the Proposition above can be improved.
If, for instance, $\nu_1$ has a bounded density, say $f_1$, then $G(\mu)$ has a bounded
density, $g$, which satisfies
$$\|g\|_\infty\leq \lambda_1^{-d} \|f_1\|_\infty.$$
To check this note that for almost every $x$ the Alexandrov Hessian $D^2\varphi(x)=\sum_{j=1}^k \lambda_j D^2 \varphi_j(x)$
can be expressed as 
$$D^2\varphi(x)=\lambda_1 D^2 \varphi_1(x)\left( Id+AB\right)$$
$A=\lambda_1^{-1}D^2 \varphi_1(x)^{-1}$, $B=\lambda_1^{-1}D^2 \varphi_1(x)^{-1}\Big( \sum_{j=2}^k \lambda_j D^2 \varphi_j(x)\Big)$.
But $\det(Id+AB)=\det(Id+A^{1/2}B A^{1/2})\geq 1$ since $A$ and $B$ are (a.s.) positive definite and, therefore, 
$$\det(D^2\varphi(x))\geq \lambda_1^{d} \det(D^{2}\varphi_1(x)).$$
On the other hand, writing $g_0$ for the density of $\mu$, by the Monge-Amp\`ere equation (see Theorem 4.8 in \cite{Villani}) we have that a.s. 
$$g(\nabla\varphi(x))=\frac{g_0(x)}{\det(D^2\varphi(x))}\leq \lambda_1^{-d}\frac{g_0(x)}{\det(D^2\varphi_1(x))}=f_1(\nabla \varphi_1(x))\leq 
\lambda_1^{-d} \|f_1\|_\infty.$$
As a consequence, $g(y)=g(\nabla\varphi(\nabla\varphi^*(y)))\leq \lambda_1^{-d} \|f_1\|_\infty$ a.s., as claimed.} 
\end{Remark}

\bigskip
Our next result provides a link between the $G$ transform and the barycenter problem.

\begin{Proposition}\label{basicineq}
If $\mu\in\mathcal{P}_{2,ac}(\Read)$ then
\begin{equation}\label{desigbasica}
V(\mu)\geq V(G(\mu))+\mathcal{W}_2^2(\mu,G(\mu)).
\end{equation}
As a consequence, $V(\mu)\geq V(G(\mu))$, with strict inequality if $\mu\ne G(\mu)$.
In particular,  if $\mu$ is a barycenter then $G(\mu)=\mu$.
\end{Proposition}

\medskip
\noindent
\textbf{Proof.} We simply note that for any $x,x_1,\ldots,x_k\in\mathbb{R}^d$, writing $\bar{x}=\sum_{j=1}^k\lambda_j x_j$,
we have $ \sum_{j=1}^k\lambda_j \|x-x_j\|^2=\sum_{j=1}^k \lambda_j\|\bar{x}-x_j\|^2+\|x-\bar{x}\|^2$.
As a consequence, writing as above $T_j$ for the optimal transportation map from $\mu$ to $\nu_j$ 
and $\bar{T}(x)=\sum_{j=1}^k\lambda_j T_j(x)$, then
\begin{eqnarray}
V(\mu)&=&\sum_{j=1}^k \lambda_j\int \|x-T_j(x) \|^2 d\mu(x)
= \int  \sum_{j=1}^k\lambda_j \|x-T_j(x) \|^2 d\mu(x)\label{eq1}\\
&=& \nonumber
\int \sum_{j=1}^k \lambda_j \|\bar{T}(x)-T_j(x) \|^2 d\mu(x)+\int 
\|{\textstyle \bar{T}(x) }-x \|^2 d\mu(x).
\end{eqnarray}
Also note that, from Remark \ref{cyclicmon}, $\bar{T}$ is an optimal transportation map from $\mu$ to $G(\mu)$ and
\begin{equation}\label{eq2}
\int \|{\textstyle \bar{T}(x) }-x \|^2 d\mu(x)=\mathcal{W}_2^2(\mu,G(\mu)).
\end{equation}
Finally, writing $\tilde{\pi}_j$ for the probability induced from $\mu$ by the map $(T_j,\bar{T})$
we see that $\tilde{\pi}_j$ is a coupling of $\nu_j$ and $G(\mu)$ and, as a consequence
\begin{eqnarray}\nonumber
\int \sum_{j=1}^k \lambda_j \|\bar{T}(x)-T_j(x) \|^2 d\mu(x)&=&
\sum_{j=1}^k \lambda_j \int \|y-x_j \|^2 d\tilde{\pi}_j(x_j,y)\\ \label{eq3}
&\geq & \sum_{j=1}^k \lambda_j \mathcal{W}_2^2(G(\mu),\nu_j)=V(G(\mu)).
\end{eqnarray}
Combining (\ref{eq1}), (\ref{eq2}) and (\ref{eq3}) we get (\ref{desigbasica}).
Obviously, this implies that
$V(\mu)\geq V(G(\mu))$, with strict inequality unless $G(\mu)=\mu$. In particular, if 
$\mu\in\mathcal{P}_{2,ac}(\Read)$ is a barycenter then the
inequality cannot be strict and, consequently, we must have $G(\mu)=\mu$.\quad $\Box$

\bigskip
 
 \begin{Remark}\label{extensionProp}
 {\em We notice that Proposition \ref{basicineq} remains true under a more general setting. Assume just that $\mu$ and $\nu_j, j=1,\dots,k$ belong to $\mathcal{P}_{2}(\Read)$ and let $\pi_j$ be optimal couplings of $\mu$ and $\nu_j$. By the gluing lemma, we can assume that $\pi_j=\mathcal L(X,Y_j)$ for random $\Read-$valued vectors $X,Y_1,\dots,Y_k$ defined in some probability space. Note that, in general, there exist multiple joint distributions of $(X,Y_1,\dots,Y_k)$ compatible with this construction. Nevertheless, for each of them, we can consider $\bar Y:=\sum_{j=1}^k\lambda_jY_j$ and the distribution $\mathcal L(X,\bar Y)$ that, as noted in Remark \ref{cyclicmon}, will be an optimal coupling of $\mu$ and $\mathcal L(\bar Y)$. Therefore, setting $\tilde G(\mu)=\mathcal L(\bar Y)$ (observe that, however, $\tilde G$ is not uniquely defined), we can replicate the argument in the proof above to obtain $V(\mu)\geq V(\tilde G(\mu))+\mathcal{W}_2^2(\mu,\tilde G(\mu))$ with strict inequality if $\mu\ne \tilde G(\mu).$
 
}
\end{Remark}

\bigskip

We include now a simple but important consequence of Proposition \ref{basicineq}, which follows after observing that, as noted in Remark 3.2, under (\ref{asum1}) the unique barycenter has a density.

\begin{Corollary}\label{corolarionuevo}

Under (\ref{asum1}), if $\bar{\mu}$ is the unique barycenter of $\nu_1,\ldots,\nu_k$ then $G(\bar{\mu})=\bar{\mu}$.

\end{Corollary}

We observe that Corollary \ref{corolarionuevo} can be obtained as a consequence of the duality results in \cite{AguehCarlier} (see, in particular, Remark 3.9 there) but deducing it from (\ref{desigbasica}) is simpler. Furthermore, Proposition 3.3 and Corollary \ref{corolarionuevo} are the basis for considering the following iterative procedure.
We start from $\mu_0\in \mathcal{P}_{2,ac}(\Read)$ and consider the sequence
\begin{equation}\label{iteration}
\mu_{n+1}:=G(\mu_n),\quad n\geq 0.
\end{equation}
By Theorem \ref{p1} the iteration is well defined. We provide now the consistency framework
of the sequence $\mu_n$.

\begin{Theorem}\label{PrevioConsistencia}
The sequence $\{\mu_n\}$ defined in (\ref{iteration}) is tight. Under (\ref{asum1}) every weakly convergent subsequence
of $\mu_n$ must converge in $\mathcal{W}_2$ distance to a probability in $\mathcal{P}_{2,ac}(\Read)$
which is a fixed point of $G$. In particular, under (\ref{asum1}), if $G$ has a unique fixed point, $\bar{\mu}$, then
$\bar{\mu}$ is the barycenter of $\nu_1,\ldots,\nu_k$ and
$\mathcal{W}_2(\mu_n,\bar{\mu})\to 0$.
\end{Theorem}

\medskip
\noindent
\textbf{Proof.} We consider a random vector $X_n$ with law $\mu_n$ and write
$T_{n,j}$ for the optimal transportation map from $\mu_n$ to $\nu_j$ and
$Y_{n,j}=T_{n,j}(X_n)$, $j=1,\ldots,k$. The sequence $\{(Y_{n,1},\ldots,Y_{n,k})\}_n$ has
fixed marginals, hence it is tight. By continuity, $\{\sum_{j=1}^k \lambda_j Y_{n,j}\}_n$ 
is also a tight sequence. But $\mu_n=\mathcal{L}(\sum_{j=1}^k \lambda_j Y_{n,j})$.
Arguing as in the proof of Theorem \ref{p1} we see that $\|\cdot\|^2$ is uniformly 
$\mu_n$-integrable. Hence, any weakly convergent subsequence of $\mu_n$
converges also in $\mathcal{W}_2$, as claimed. 

Assume now that (\ref{asum1}) holds. Without loss of generality we assume that $\nu_1$ has a bounded density, $f_1$.
From Remark \ref{boundeddensity} we have that $\nu_n(A)\leq \lambda_1^{-d} \|f_1\|_\infty \ell_d(A)$ for every Borel $A$. Hence, if $\tilde{\mu}$
is a weak limit of a subsequence $\mu_{n_m}$ then $\tilde{\mu}(A)\leq \liminf_{m\to\infty} \mu_{n_m}(A)\leq 
\lambda_1^{-d} \|f_1\|_\infty \ell_d(A)$ for every
open $A$ and, as a consequence, $\tilde{\mu}$ has a density (upper bounded by $\lambda_1^{-d} \|f_1\|_\infty$).
Since we have $\mathcal{W}_2(\mu_{n_m},\tilde{\mu})\to 0$
as $m\to\infty$, by the continuity result in Theorem
\ref{p1} we have $\mathcal{W}_2(\mu_{n_m+1},G(\tilde{\mu}))\to 0$ as well.
Now, $V$ is continuous in $\mathcal{W}_2$ metric (this follows, for instance, 
from the fact that $|V(\mu_1)^{1/2}-V^{1/2}(\mu_2)|\leq \mathcal{W}_2(\mu_1,\mu_2)$).
As a consequence, $V(\mu_{n_m})\to V(\tilde{\mu})$ and 
$V(\mu_{n_m+1})\to V(G(\tilde{\mu}))$. But by Proposition \ref{basicineq}
$V(\mu_n)$ is a nonnegative and nonincreasing sequence, hence, it is convergent. Therefore,
$$V(\tilde{\mu})=\lim_{m\to\infty} V(\mu_{n_m})=\lim_{m\to\infty} V(\mu_{n_m+1})=V(G(\tilde{\mu})).$$
Using again Theorem \ref{p1} we see that $\tilde{\mu}$ must satisfy $G(\tilde{\mu})=\tilde{\mu}$.
Under assumption (\ref{asum1}) the (unique) barycenter, $\bar{\mu}$, is a fixed point
for $G$. If $G$ has a unique fixed point then every subsequence of $\mu_n$
has a further subsequence that converges to $\bar{\mu}$ in $\mathcal{W}_2$. Hence,
$\mathcal{W}_2(\mu_n,\bar{\mu})\to 0$. \quad $\Box$

\bigskip
In some special cases we can guarantee uniqueness of the fixed point and the above result can be sharpened as we show in the next section. Observe that the fixed point condition $G(\mu)=\mu, \ \mu \in \mathcal{P}_{2,ac}(\Read)$ is equivalent to $\sum_{j=1}^k\lambda_jT_j(x)=x$ $\mu-$a.s. If this equality holds for every $x \in \Read$ then $\mu$ is a barycenter (see Remark 3.9 in \cite{AguehCarlier}). However, this is not always the case as the following example shows. Providing sufficient conditions on $\nu_1,\ldots,\nu_k$ to guarantee that $G$ has a unique fixed point is a goal for future work.

\begin{Example}
{\rm Consider the set ${\cal S}=\{A,B,C,D\}$, where  $A=(0,1)$,  $B=(1,1)$,  $C=(1,0)$,  and  $D=(0,0)$. Given $ S\in {\cal S}$, let $\Prob_S$ be the uniform distribution on the three points ${\cal S}-\{S\}$, and let $\mu_1$ be the uniform discrete distribution, supported on 12 points uniformly scattered on the four sides of the unit square excluding the vertices:
\[
P_1=(.25,1), P_2=(.5,1), P_3=(.75,1), P_4=(1,.75), \ldots, P_{11}=(0,.5), P_{12}= (0,.75).
\]

It is very easy to prove that the maps $T_S, S\in {\cal S}$,  described in Table \ref{Tabla.1}, are the only optimal transport maps between $\mu_1$ and the probabilities $\Prob_S,  S\in {\cal S}$ respectively.

\begin{table}[ht]
\begin{center}
\begin{tabular}{l|cccc}
Map & A & B & C & D
\\
\hline
$T_A$ & - & $P_1$, $P_2$, $P_3$, $P_4$, & $P_5$, $P_6$, $P_7$, $P_8$ & $P_9$, $P_{10}$, $P_{11}$, $P_{12}$
\\
$T_B$ & $P_{12}$, $P_1$, $P_2$, $P_3$ & -  & $P_4$, $P_5$, $P_6$, $P_7$ & $P_8$, $P_9 $, $P_{10} $, $P_{11}$
\\
$T_C$ & $P_{11}$, $P_{12}$, $P_1$, $P_2$ & $P_3$, $P_4$, $P_5$, $P_6$ & - & $P_7$, $P_8$, $P_9$, $P_{10}$
\\
$T_D$ & $P_{10}$, $P_{11}$, $P_{12}$, $P_ 1$ & $P_2$, $P_3$, $P_4$, $P_5$ & $P_6$, $P_7$, $P_8$, $P_9$ & - 
\\
\hline
\end{tabular} 
\caption{Points going to each point in $\cal S$ for the four optimal transports  between $\mu_1$ and the $\Prob_S$'s}\label{Tabla.1}
\end{center}
\end{table}

Given $\delta, \gamma>0$, let us consider the distributions $\nu_S^{\delta}$ which are uniform on the union of the three balls with radius $\delta$ and centers in  ${\cal S}-\{S\}$, and  $\mu_1^\gamma$ uniform on the union of the twelve balls with centers at $P_1,\ldots, P_{12}$ and radius $\gamma$. Let $T_S^{\gamma,\delta}$ denote the optimal transport map between $\mu_1^\gamma$ and $\nu_S^\delta$. If we take $\delta, \gamma>0$ small enough, 
 then it is obvious that $T_S^{\gamma,\delta} (B(P_i,\gamma))\subset B(T_S(P_i),\delta)$ for every $i=1,\dots,12$ and
 also  that  we can choose $\gamma >\delta$ in such a way that, for every $x \in \Rea^2$, the ball $B(x,\gamma)$ contains the square with side $\delta$ and center at $x$.

Now, let us consider the map
\[
x \to  \overline{T}^{\gamma,\delta}(x) = \frac 1 4 \sum_{S \in {\cal S}} T_S^{\gamma,\delta}(x)=:(t_1,t_2).
\]

Assume, for instance, that $x\in B(P_1,\gamma)$. Then, according to Table \ref{Tabla.1}, we have that $T_A^{\gamma,\delta}(x) \in B(B,\delta)$ and $T_S^{\gamma,\delta}(x) \in B(A,\delta)$, for every $S=B,C,D$, hence, 
 we have that
\begin{eqnarray*}
.25-\delta \leq & t_1 & \leq .25+\delta
\\
1-\delta \leq & t_2 &  \leq1+\delta.
\end{eqnarray*}
Therefore $\overline{T}^{\gamma,\delta}(x)$ belongs to the square with side $\delta$ and center at $P_1$, and, consequently, $\overline{T}^{\gamma,\delta}(B(P_1,\gamma)) \subset B(P_1,\gamma)$

Obviously, the same happens for every point in the support of $\mu_1^\gamma$ and we can conclude that $\overline{T}^{\gamma,\delta}(B(P_i,\gamma)) \subset B(P_i,\gamma), i=1,\ldots, 12$. Moreover, we can iterate the procedure and define  $\mu_1^*$ to be a limit point (through some convergent subsequence) of the process $G_\delta( \cdots G_\delta(\mu_1^\gamma))$, where $G_\delta$ is the operator associated to the family $\{\nu_S^\delta, S\in {\cal S}\}$ with uniform weights. According to Theorem \ref{PrevioConsistencia} $\mu_1^*\in  \mathcal{P}_{2,ac}(\Read)$ and it is a fixed point of $G_\delta$ that, by the previous argument,  satisfies $\mu_1^*(B(P_i,\delta))=\mu_1(B(P_i,\delta))$,  for every $i=1,\ldots,12$.

To get a second fixed point of $G_\delta$, let us consider the probability $\mu_2$ supported by the points $P_1,\ldots, P_{13}$, where $P_1,\ldots, P_{12}$ are as before and $P_{13}=(.5,.5)$. Let  $l=3^{-1/2}$ and $p=.5*(1-l)^2, \ q= (2*l-1)*(1-l), \ r=(2*l-1)^2,$
and define
\[
\mu_2(P_i) =
\left\{
\begin{array}{ll}
p,  \mbox{ if }  i=1, 3, 4, 6, 7, 9, 10, 12
\\
q,  \mbox{ if }  i=2, 5, 8, 11
\\
r,  \mbox{ if } i=13
\end{array}
\right.
\]

It can be checked that $8p + 4q + r =1$, thus $\mu_2$ is a probability distribution. As in the previous case, the optimal transportation maps, $T_S^*, S\in {\cal S}$,  between $\mu_2$ and the probabilities $\Prob_S,  S\in {\cal S}$  are described in Table \ref{Tabla2}.

\begin{table}[ht]
\begin{center}
\begin{tabular}{l|cccc}
Map & A & B & C & D
\\
\hline
$T_A^*$ & - & $P_1$, $P_2$, $P_3$, $P_4$, & $P_5$, $P_6$, $P_7$, $P_8$, $P_{13}$ & $P_9$, $P_{10}$, $P_{11}$, $P_{12}$
\\
$T_B^*$ & $P_{12}$, $P_1$, $P_2$, $P_3$ & -  & $P_4$, $P_5$, $P_6$, $P_7$ & $P_8$, $P_9 $, $P_{10} $, $P_{11}$, $P_{13}$
\\
$T_C^*$ & $P_{11}$, $P_{12}$, $P_1$, $P_2$, $P_{13}$ & $P_3$, $P_4$, $P_5$, $P_6$ & - & $P_7$, $P_8$, $P_9$, $P_{10}$
\\
$T_D^*$ & $P_{10}$, $P_{11}$, $P_{12}$, $P_ 1$ & $P_2$, $P_3$, $P_4$, $P_5$, $P_{13}$ & $P_6$, $P_7$, $P_8$, $P_9$ & - 
\\
\hline
\end{tabular} 
\caption{Points going to each point in $\cal S$ for the four optimal transports between $\mu_2$ and the $\Prob_S$'s}\label{Tabla2}
\end{center}
\end{table}

From this point on, it is possible to repeat the construction, leading to $\mu_1^*$, to obtain from $\mu_2$ and suitable values $\delta,\gamma$ an absolutely continuous $\mu_2^*$ (verifying $\mu_2^*(B(P_i,\delta))=\mu_2(B(P_i,\delta)), i=1,\dots,13$) which is also a fixed point of $G_\delta$. If we take the smaller values obtained for $\delta$ and $\gamma$ both constructions work and, then, we have obtained two different fixed points for $G_\delta$ since $\mu_1^* \neq \mu^*_2$.

}
\end{Example}

\section{Barycenters in location-scatter families.}\label{cuatro}

We focus now on the barycenter problem in the special 
case in which $\nu_1,\ldots,\nu_k$ are probabilities in the same location-scatter
family
$$\mathcal{F}(P_0):=\{\mathcal{L}(A{\bf X}_0+m): A\in {\cal M}_{d\times d}^+, m\in\Read\},$$
where, as before, $ {\cal M}_{d\times d}^+$ denotes the set of $d\times d$ positive definite matrices  
and is $\bf{X}_0$ a random vector with probability law  $P_0$. Also, as in Section 2, and
without loss of generality, we will assume that $P_0$ is centered and has the identity as covariance matrix.
We will provide a consistent iterative method for the computation of the barycenter of $\nu_1,\ldots,\nu_k\in 
\mathcal{F}(P_0)$.
We remark that this covers the case when
$\nu_1,\ldots,\nu_k$ are Gaussian or belong to the same elliptical family (but it is not constrained
to these cases: if we take in $\mathbb{R}^2$ the probability $P_0$ whose marginals are independent 
standard uniform laws, or even standard normal and exponential, then the family $\mathcal{F}(P_0)$ is not elliptical).
In particular, our approach will give a self-contained proof of the fact that equation (\ref{matrixeq}) 
has a unique symmetric, positive definite solution.

\medskip
Let us focus first in the case when $\nu_1,\ldots,\nu_k$ are (nondegenerate) centered Gaussians, say $\nu_j=N(0,\Sigma_j)$, $j=1,\ldots,k$.
We know that there exists a unique barycenter. On the other hand,
since $\sum_{j=1}^k\lambda_j \|\bar{x}-x_j\|^2=\sum_{j=1}^k\lambda_j \|x_j\|^2-\|\bar{x}\|^2$, we see that a minimizer in the 
multimarginal formulation (recall the discussion about existence and uniqueness of barycenters after Definition \ref{baricentro}) is the 
law of a random vector $(X_1,\ldots,X_k)$ with $\mathcal{L}(X_j)=N(0,\Sigma_j)$ that maximizes
$E\|\lambda_1 X_1+\cdots+\lambda_k X_k\|^2=\sum_j \lambda_j^2 E\|X_j\|^2+ 2\sum_{1\leq j<l\leq k} \lambda_j\lambda_jE(X_j\cdot X_l)$. 
From this last expression we see that $E\|\lambda_1 X_1+\cdots+\lambda_k X_k\|^2$ depends only on the covariance structure
of  the $k\times d$-dimensional vector $(X_1,\ldots,X_k)$. Given any covariance structure we can find a centered Gaussian random vector with that covariance structure.
Hence, a centered Gaussian minimizer exists for the multimarginal problem and, by linearity of $T$, a centered Gaussian barycenter
exists for $\nu_1,\ldots,\nu_k$. By uniqueness of the barycenter, this shows that \textit{the barycenter of nondegenerate centered
Gaussian distributions is a nondegenerate centered Gaussian distribution}.

\medskip
With this fact in mind, and abusing notation, we write  $V(\Sigma)$ for $V(N(0,\Sigma))$  and consider the problem of minimizing
$V(\Sigma)$. Note that
\begin{equation}\label{Vfunction}
V(\Sigma)=\mbox{Tr}(\Sigma)+\sum_{j=1}^k \lambda_j \mbox{Tr}(\Sigma_j)-2\sum_{j=1}^k \lambda_j\mbox{Tr}\big( (\Sigma^{1/2}\Sigma_j\Sigma^{1/2})^{1/2}\big).
\end{equation}
We also write $G$
for the map $\Sigma\mapsto \Sigma^{-1/2}\Big(\sum_{j=1}^k \lambda_j (\Sigma^{1/2}\Sigma_j\Sigma^{1/2})^{1/2}\Big)^2 \Sigma^{-1/2}$.
With this notation we have the following upper and lower bounds.
\begin{Proposition}\label{upperlowerbounds}
With the above notation, if $\Sigma \in  {\cal M}_{d\times d}^+$:
\begin{eqnarray}\label{desig1normal}
V(\Sigma)-V(G(\Sigma))
&\geq &
\mbox{\em Tr}\big(\Sigma^{1/2}(Id-H(\Sigma))^2\Sigma^{1/2}\big),\quad \mbox{and for any } \Sigma'  \in {\cal M}_{d\times d}^+
\\
\label{desig2normal}
V(\Sigma')-V(\Sigma)
&\geq &
\mbox{\em Tr}((Id-H(\Sigma))(\Sigma'-\Sigma)),
\end{eqnarray}
where $H(\Sigma)=\sum_{j=1}^k\lambda_j \Sigma^{-1/2} (\Sigma^{1/2}\Sigma_j\Sigma^{1/2})^{1/2}\Sigma^{-1/2}$.
\end{Proposition}

\medskip
\noindent \textbf{Proof.} We note first that (\ref{desig1normal}) is just the particular form of (\ref{desigbasica}) 
in the present setup. For the other bound, denoting by $H_j(\Sigma)$  the matrix associated to the optimal transportation map from $N(0,\Sigma)$ to
$N(0,\Sigma_j)$, namely, $H_j(\Sigma)=\Sigma^{-1/2} (\Sigma^{1/2}\Sigma_j\Sigma^{1/2})^{1/2}\Sigma^{-1/2}$, from
the fact that for any $A \in  {\cal M}_{d\times d}^+$ we have  $x^TtA x+y^t A^{-1}y\geq 2 x\cdot y$
for every $x,y$, with equality if and only if $y=A x$,
we see that for any $(\mathbb{R}^d)^2$-valued random vector $(X,X_j)$ such that $X\sim N(0,\Sigma')$ and $X_j\sim N(0,\Sigma_j)$,
\begin{eqnarray}\nonumber
E\|X-X_j\|^2&\geq &
E(\|X\|^2-X^tH_j(\Sigma)X)+E(\|X_j\|^2-X_j^tH_j(\Sigma)^{-1}X_j)\\\label{lowerbound}
&=&\mbox{Tr}((Id-H_j(\Sigma))\Sigma')+E(\|X_j\|^2-X_j^tH_j(\Sigma)^{-1}X_j).
\end{eqnarray}
From this we conclude that
$$V(\Sigma')\geq \mbox{Tr}((Id-H(\Sigma))\Sigma')+\sum_{j=1}^k \lambda_j E(\|X_j\|^2-X_j^tH_j(\Sigma)^{-1}X_j).$$
In the case $\Sigma'=\Sigma$, if $X\sim N(0,\Sigma)$ and $X_j=H_j(\Sigma) \circ X$ then (\ref{lowerbound})
becomes an equality and we get
$$\mathcal{W}_2^2(N(0,\Sigma),N(0,\Sigma_j))=E\|X-X_j\|^2=\mbox{Tr}((Id-H_j(\Sigma))\Sigma)+E(\|X_j\|^2-X_j^tH_j(\Sigma)^{-1}X_j),$$
from which we obtain (\ref{desig2normal}).\quad $\Box$

\medskip
Observe now that for any $\Sigma \in  {\cal M}_{d\times d}^+$, (\ref{desig1normal}) shows that $H(\Sigma)=Id$
is a necessary condition for $N(0,\Sigma)$ to be a barycenter, while (\ref{desig2normal}) shows that
the condition is sufficient as well. Thus, recalling the discussion before Proposition \ref{upperlowerbounds},
if we knew that the barycenter of nondegenerate centered
Gaussian distributions is a centered, \textit{nondegenerate} Gaussian distribution, then, we could conclude
that it must be $N(0,\Sigma)$, with $\Sigma$ the unique solution to the matrix equation (\ref{matrixeq}), which, of course, is equivalent
to the condition $H(\Sigma)=Id$. Through the consideration of the iteration introduced in Section 3 we will show next that, indeed, 
there exists $\Sigma \in  {\cal M}_{d\times d}^+$ such that 
$H(\Sigma)=Id$. This will yield an alternative proof of Theorem \ref{casonormal}. Furthermore, it will provide
a consistent iterative method for the approximation of Gaussian barycenters.
\begin{Theorem}\label{normaliteration}
Assume $\Sigma_1,\ldots,\Sigma_k$ are symmetric $d\times d$ positive semidefinite matrices, with at least one of them positive definite.
Consider some $S_0 \in  {\cal M}_{d\times d}^+$ and define
\begin{equation}\label{iterationnormal}
S_{n+1}=S_{n}^{-1/2}\Big(\sum_{j=1}^k \lambda_j(S_{n}^{1/2}\Sigma_j S_{n}^{1/2})^{1/2} \Big)^2 S_{n}^{-1/2},\quad n\geq 0.
\end{equation}

If $N(0,\Sigma_0)$
is the barycenter of $N(0,\Sigma_1),\ldots,N(0,\Sigma_k)$, then 
$$
\mathcal{W}_2(N(0,S_{n}),N(0,\Sigma_0))\to 0
$$ 
as $n\to \infty$. Moreover, the covariance matrix
of the barycenter satisfies
\begin{equation}\label{lowerbounddetbarb}
\det(\Sigma_0)^{1/2d}\geq \sum_{j=1}^k \lambda_j \big(\det(\Sigma_j)\big)^{1/2d}.
\end{equation}
In particular, it is positive definite and it is the unique positive definite solution to 
(\ref{matrixeq}). Furthermore, 
\begin{equation}\label{upperboundtrabarb}
\mbox{\em Tr}(S_{n})\leq \mbox{\em Tr}(S_{n+1}) \leq \mbox{\em Tr}(\Sigma_0)\leq \sum_{j=1}^k \lambda_j \mbox{\em Tr}(\Sigma_j),
\end{equation}
with equality in last inequality if and only if $\Sigma_1=\cdots=\Sigma_k$.

\end{Theorem}

\medskip
\noindent
\textbf{Proof.} Let us begin proving that for $n\geq 1$,
\begin{equation}\label{lowerbounddetbar}
\det(S_{n})^{1/2d}\geq \sum_{j=1}^k \lambda_j \big(\det(\Sigma_j)\big)^{1/2d},
\end{equation}
which, in particular,  gives that  all the $S_n$ are nonsingular and the sequence is well defined.
To this  just note that by
the Minkowski determinant inequality (see, e.g., Corollary II.3.21 in \cite{Bhatia}) we get
\begin{eqnarray*}
\Big(\det\big(\sum_{j=1}^k \lambda_k (S_n^{1/2}\Sigma_j S_n^{1/2})^{1/2} \Big)\Big)^{1/d}
&\geq & \sum_{j=1}^k \lambda_j \big(\det(S_n^{1/2}\Sigma_j S_n^{1/2})\big)^{1/2d}\\
&=&\det(S_n)^{1/2d}\sum_{j=1}^k \lambda_j \big(\det(\Sigma_j)\big)^{1/2d},
\end{eqnarray*}
from which (\ref{lowerbounddetbar}) follows.
Tightness of the sequence
$N(0,S_n)$ (which follows from Theorem \ref{PrevioConsistencia}) implies boundedness of $\{S_n\}$. 
We take a convergent subsequence $S_{n_m}\to \Sigma$. By continuity we have
$\det(\Sigma)^{1/2d}\geq \sum_{j=1}^k\lambda_j \det(\Sigma_j)^{1/2d}>0$, which shows that $\Sigma \in  {\cal M}_{d\times d}^+.$. 
The map $V$ is continuous on the set $ {\cal M}_{d\times d}^+$ and, therefore, $V(S_{n_m})\to V(\Sigma)$. Continuity of the map $G$ implies that
$V(S_{n_m+1})\to V(G(\Sigma))$. Finally, the fact that
$V(S_n)$ is a nonnegative, nonincreasing sequence implies that it is convergent. Hence, we have
$$V(\Sigma)=V(G(\Sigma)).$$

In view of (\ref{desig1normal}), this can only happen if $\Sigma^{1/2}(Id-H(\Sigma))^2\Sigma^{1/2}=0$. Since $\Sigma \in {\cal M}_{d\times d}^+$ 
this can happen only if $Id-H(\Sigma)=0$. Hence, we have proved that equation (\ref{matrixeq}) has a unique positive definite 
solution which corresponds to the covariance matrix of the barycenter of $\nu_1,\ldots,\nu_k$. Thus $\Sigma=\Sigma_0$. 
Using again Theorem \ref{PrevioConsistencia} we conclude that $\mathcal{W}_2(N(0,S_{n}),N(0,\Sigma_0))\to 0$ as $n\to \infty$.
By continuity, (\ref{lowerbounddetbarb}) follows.

For the first inequality in (\ref{upperboundtrabarb}) we write $H_n$ for the matrix of the optimal transportation map from
$N(0,S_n)$, to $N(0,S_{n+1})$, namely, $H_n=S_n^{-1/2}\sum_{j=1}^k \lambda_j (S_n^{1/2}\Sigma_j S_n^{1/2})^{1/2}S_n^{-1/2}$.
Note that $S_{n+1}=H_n S_n H_n$ and, as a consequence, $(S_{n}^{1/2}S_{n+1}S_{n}^{1/2})^{1/2}=S_{n}^{1/2}H_nS_{n}^{1/2}$, from
which we conclude that
\begin{equation}\label{expre1}
\mathcal{W}_2^2(N(0,S_n),N(0,S_{n+1}))=\mbox{Tr}(S_n)+\mbox{Tr}(S_{n+1})-2\mbox{Tr}(S_nH_n).
\end{equation}

On the other hand, 
$$V(S_n)=\mbox{Tr}(S_n)+\sum_{j=1}^k \lambda_j \mbox{Tr}(\Sigma_j)-2 \mbox{Tr}(S_nH_n), $$
which shows that
\begin{equation}\label{expre2}
V(S_n)-V(S_{n+1})=(\mbox{Tr}(S_n)-2\mbox{Tr}(S_nH_n))-(\mbox{Tr}(S_{n+1})-2 \mbox{Tr}(S_{n+1}H_{n+1})).
\end{equation}

Now combining the lower bound (\ref{desigbasica}) with (\ref{expre1}) and (\ref{expre2}) we obtain
\begin{eqnarray*}
V(S_n)-V(S_{n+1})&=&(\mbox{Tr}(S_n)-2\mbox{Tr}(S_nH_n))-(\mbox{Tr}(S_{n+1})-\mbox{Tr}(S_{n+1}H_{n+1}))\\
&\geq & \mathcal{W}_2^2(N(0,S_n),N(0,S_{n+1}))=\mbox{Tr}(S_n)+\mbox{Tr}(S_{n+1})-2\mbox{Tr}(S_nH_n),
\end{eqnarray*}
which entails $2 (\mbox{Tr}(S_{n+1} H_{n+1})-\mbox{Tr}(S_{n+1}))\geq 0$ and, therefore,
\begin{equation}\label{bnd1}
\mbox{Tr}(S_{n+1})\leq\mbox{Tr}(S_{n+1} H_{n+1}).
\end{equation}

Moreover, since
$$0\leq \mathcal{W}_2^2(N(0,S_{n+1}),N(0,S_{n+2}))=(\mbox{Tr}(S_{n+1})-\mbox{Tr}(S_{n+1} H_{n+1}))+(\mbox{Tr}(S_{n+2})-\mbox{Tr}(S_{n+1} H_{n+1})),$$
we must have 
$$\mbox{Tr}(S_{n+1} H_{n+1})\leq \mbox{Tr}(S_{n+2}).$$

This and (\ref{bnd1}) prove the first inequality in (\ref{upperboundtrabarb}). To show that $\mbox{\em Tr}(S_{n})\leq  \sum_{j=1}^k \lambda_j \mbox{\em Tr}(\Sigma_j) $, note that
$$0\leq V(S_{n+1})=(\mbox{Tr}(S_{n+1})-\mbox{Tr}(S_{n+1} H_{n+1}))+\big(\sum_{j=1}^k \lambda_k \mbox{Tr}(\Sigma_j)-\mbox{Tr}(S_{n+1} H_{n+1})\big)$$
and, therefore,
$$\mbox{Tr}(S_{n+1})\leq \mbox{Tr}(S_{n+1} H_{n+1})\leq \sum_{j=1}^k \lambda_k \mbox{Tr}(\Sigma_j).$$
Last inequality in (\ref{upperboundtrabarb}) follows by continuity. Alternatively, using that 
$\mbox{Tr}(\Sigma_0)=\sum_{j=1}^k \lambda_k \mbox{Tr}\big((\Sigma_0^{1/2}\Sigma_j \Sigma_0^{1/2})^{1/2}\big)$
together with (\ref{Vfunction}) we see that
$$0\leq V(\Sigma_0)=\sum_{j=1}^k \lambda_j \mbox{Tr}(\Sigma_j)-\mbox{Tr}(\Sigma_0),$$
which allows us to conclude (\ref{upperboundtrabarb})
with equality if and only if $\Sigma_1=\cdots=\Sigma_k$.\quad $\Box$

\begin{Remark}{\em The already noted fact that convergence in $\mathcal{W}_2$ is equivalent to
weak convergence plus convergence of second moments shows that 
$\mathcal{W}_2(N(0,S_{n}),N(0,\Sigma_0))\to 0$ if and only if $\|S_n-\Sigma_0\|\to 0$
for some (hence, for any) matrix norm.
} 
\end{Remark}

\begin{Remark}\label{onestep}{\em In some cases iterations converge in just one step
to the barycenter. This is the case in dimension $d=1$, where $\Sigma_0=\left(\sum_{j=1}^k \lambda_j \Sigma_j^{1/2}\right)^2$
(note that in this case the lower bound (\ref{lowerbounddetbarb}) becomes an equality). More generally,
if we have $\Sigma_i \Sigma_j=\Sigma_j \Sigma_i$ for all $i,j$, then $\Sigma_{i}=U \Lambda_i U^t$, $i=1,\ldots,k$
for some orthogonal $U$ and diagonal $\Lambda_i$. It is easy to check that in this case
$$\Sigma_0=\Big(\sum_{j=1}^k \lambda_j \Sigma_j^{1/2}\Big)^2.$$
If we start the iteration from $S_0=Id$, then $S_1=\Sigma_0$ and, again, we achieve convergence in just one step.
} 
\end{Remark}

As noted before, Theorem \ref{normaliteration} yields an alternative proof of the already known fact (see 
Theorem 6.1 in \cite{AguehCarlier}) that, given  $\Sigma_1,\ldots,\Sigma_k \in {\cal M}_{d\times d}^+$ there exists a unique
 $\bar{S} \in {\cal M}_{d\times d}^+$ such that
\begin{equation}\label{matrixeqb}
\bar{S}=\sum_{j=1}^k \lambda_j (\bar{S}^{1/2}\Sigma_j \bar{S}^{1/2})^{1/2}.
\end{equation}

While this matrix equation is deeply connected to the barycenter problem, we could
read Theorem \ref{normaliteration} just in terms of approximating the unique solution
of (\ref{matrixeqb}). The conclusion becomes that if, starting from any 
$S_0 \in {\cal M}_{d\times d}^+$, we define a sequence of matrices $\{S_n\}_n$ as in (\ref{iterationnormal}),
 then 
$$\lim_{n\to\infty} S_n=\bar{S}.$$
Hence, we have provided a consistent iterative method for approximating
the solution of (\ref{matrixeq}).

\medskip
With this in mind, let us consider now probabilities in a general location-scatter family,
$\nu_j=\mathbb{P}_{m_j\Sigma_j}\in\mathcal{F}({P}_0)$, $j=1,\ldots,k$,  and recall that we are assuming
that ${P}_0$  has a density and it is centered, with $Id$ as covariance matrix. We focus on the case $m_j=0$ 
(from (\ref{centering}) we see that the barycenter in the general case equals the barycenter
corresponding to the centered probabilities  shifted by $\sum_{j=1}^k\lambda_j m_j$).
Now, the discussion at the beginning of this section applies. The minimizer in the 
multimarginal formulation is the 
law of a random vector $(X_1,\ldots,X_k)$ with $\mathcal{L}(X_j)=\nu_j$ that maximizes
$E\|\lambda_1 X_1+\cdots+\lambda_k X_k\|^2$ and this depends only on the covariance structure
of $(X_1,\ldots,X_k)$. Furthermore, we have seen that among all possible joint covariance matrices, the one that maximizes 
$E\|\lambda_1 X_1+\cdots+\lambda_k X_k\|^2$ is the covariance of the random vector $(T_1(X_0),\ldots,T_k(X_0))$
where $X_0\sim  \mathbb{P}_{0,\Sigma_0}$, $T_j(X_0)=\Sigma_0^{-1/2}(\Sigma_0^{1/2}\Sigma_j \Sigma_0^{1/2})^{1/2} \Sigma_0^{-1/2}X_0$
and $\Sigma_0$ is the unique positive definite solution to the matrix equation (\ref{matrixeqb}).
Hence, the barycenter of $\mathbb{P}_{0,\Sigma_1},\ldots,\mathbb{P}_{0,\Sigma_k}$ is the law of
$\sum_{j=1}^k\lambda_j T_j(X_0)=X_0$, that is, $\mathbb{P}_{0,\Sigma_0}$. Thus we have proved the following
consequence of Theorem \ref{normaliteration}.

\begin{Corollary}\label{locationscatter} If $\nu_j=\mathbb{P}_{m_j,\Sigma_j}\in\mathcal{F}({P}_0)$, $j=1,\ldots,k$, where 
${P}_0$ is has a density and is centered, with $Id$ as covariance matrix, then the barycenter of $\nu_1,\ldots,\nu_k$
is $\mathbb{P}_{m_0,\Sigma_0}$ with $m_0=\sum_{j=1}^k \lambda_j m_j$ and $\Sigma_0$ the unique 
positive solution of (\ref{matrixeq}). Furthermore, if starting from some positive definite $S_0$ we define $S_n$ as in
(\ref{iterationnormal}) then $\|S_n-\Sigma_0\|\to 0$ and the bounds (\ref{lowerbounddetbar}) to (\ref{upperboundtrabarb})
hold. 
\end{Corollary}

To conclude this section we observe that Corollary \ref{locationscatter} applies, for instance,
to the case of uniform distribution over ellipsoids. 
This corresponds to the case where ${P}_0$ is the uniform distribution over the ball centered at the origin with radius
$\sqrt{d+2}$ in $\mathbb{R}^d$ (a simple computation shows that $P_0$ is then centered with 
$Id$ as covariance matrix). Now, $\mathbb{P}_{m,\Sigma}$ is the uniform distribution over the ellipsoid
$E_d(m,\Sigma)=\{x\in\mathbb{R}^d:\, (x-m)^t\Sigma^{-1}(x-m)\leq d+2 \}$ and Corollary \ref{locationscatter} 
admits the following interpretation. 
On the   family of compact convex subsets of $\mathbb{R}^d$ with nonvoid interior we could consider the metric
$$w(C,D)=\mathcal{W}_2 (U_C,U_D),$$
where $U_C$ denotes the uniform distribution on $C$. Given 
$C_1,\ldots,C_k$ in that family we could consider the barycenter, namely, the compact convex set that minimizes 
$$\sum_{j=1}^{k} \lambda_k w^2(C,C_j),$$
provided it exists. In this setup, Corollary \ref{locationscatter} implies that
the barycenter of a finite collection of ellipsoids is also an ellipsoid.
More precisely, the barycenter of the ellipsoids 
$E_d(m_1,\Sigma_1),\ldots,E_d(m_k,\Sigma_k)$ is the ellipsoid
$E_d(m_0,\Sigma_0)$ with $m_0=\sum_{j=1}^k \lambda_j m_j$ and $\Sigma_0$ the unique positive definite solution to
(\ref{matrixeq}). Furthermore, Corollary \ref{locationscatter} provides an iterative method for the computation
of this barycentric ellipsoid.

 Some results in this line,  appear in \cite{Cuesta03} when the sets are required to be convex.

\section{Numerical results.}

To illustrate the performance of the iteration (\ref{iterationnormal}) we have considered the computation
of the Wasserstein barycenters in several different setups. We have applied the iterative procedure
until the difference $V(S_{n})-V(S_{n+1})$ becomes smaller than a fixed tolerance ($10^{-10}$ in all the numerical experiments
below). We recall that 
$V(S_{n})-V(S_{n+1})$ is an upper bound for the squared Wasserstein distance $\mathcal{W}_2^2(N(0,S_n),N(0,S_{n+1}))$.

We have included a comparison to the alternative 
iterative procedure considered in \cite{Ruschen2}, which can be summarized as follows. As in Theorem
\ref{normaliteration}, we 
assume  that $\Sigma_1,\ldots,\Sigma_k$ are symmetric $d\times d$ positive semidefinite matrices, with at least one of them positive definite.
We consider $S_0 \in {\cal M}_{d\times d}^+$ and compute, for $n\geq 1$
\begin{equation}\label{iterationRU}
S_{n+1}=\sum_{j=1}^k \lambda_j(S_{n}^{1/2}\Sigma_j S_{n}^{1/2})^{1/2}.
\end{equation}

While there is no theoretical consistency result for iteration (\ref{iterationRU}), in \cite{Ruschen2}
it is noted that the procedure seems to work reasonably well in the case of $k=3$ for bivariate Gaussian distributions. However ``for dimension d=3 only for favorable initial matrices convergence is observed and one has to use specific methods of numerical analysis for the solution of this nonlinear equation" (sic).

\begin{figure}[ht]\label{casobivariante}
\centerline{\includegraphics[scale=0.37]{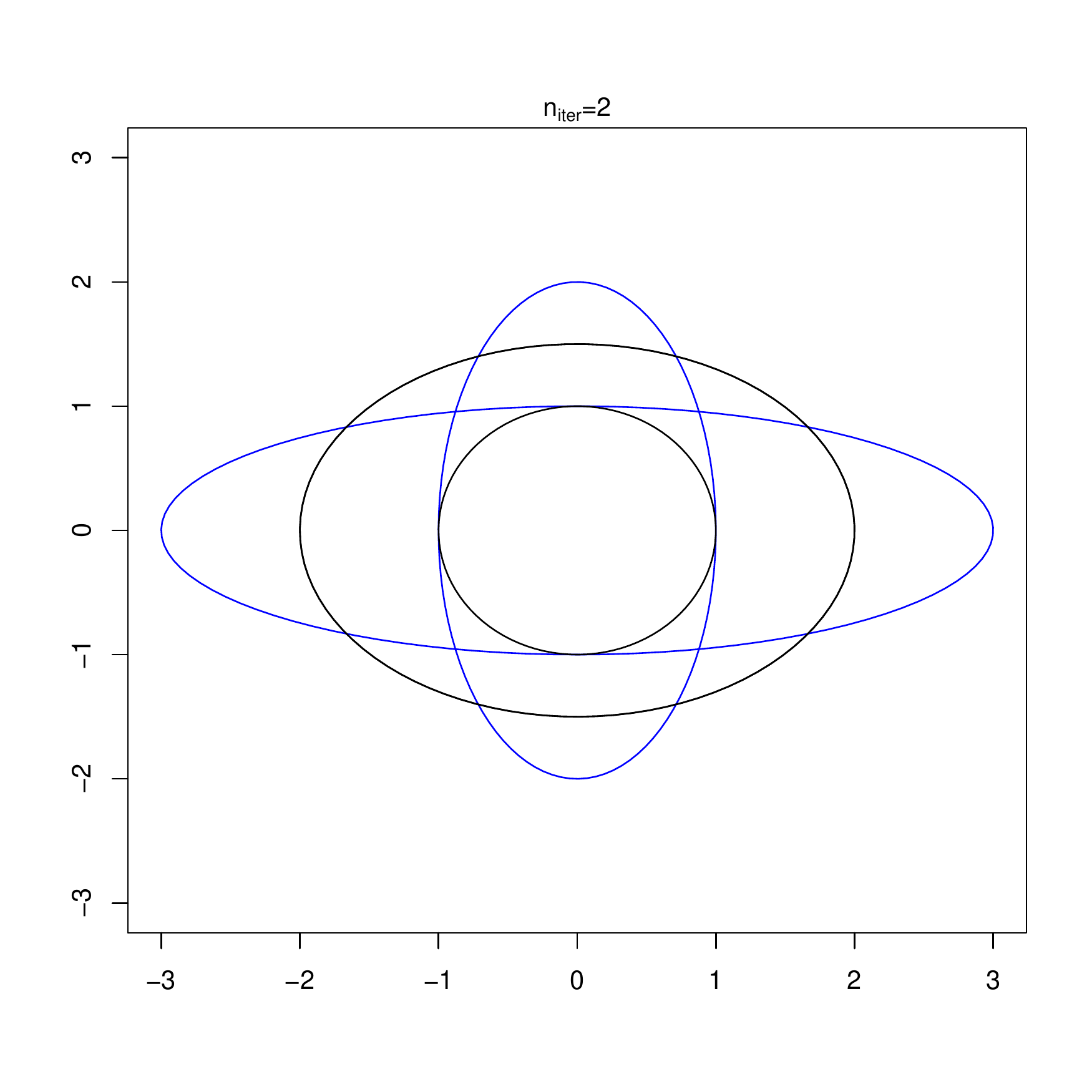}\includegraphics[scale=0.37]{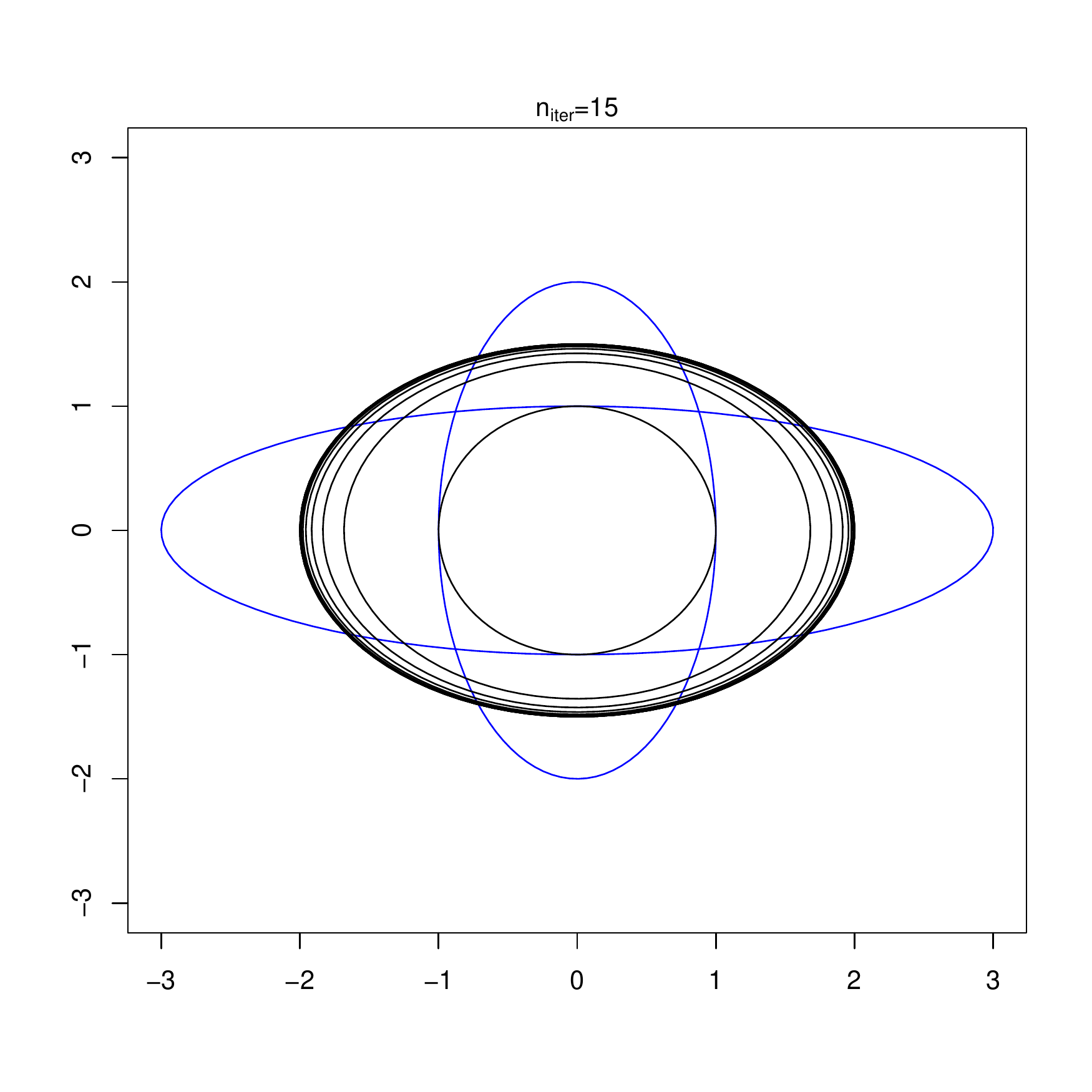}}
\centerline{\includegraphics[scale=0.37]{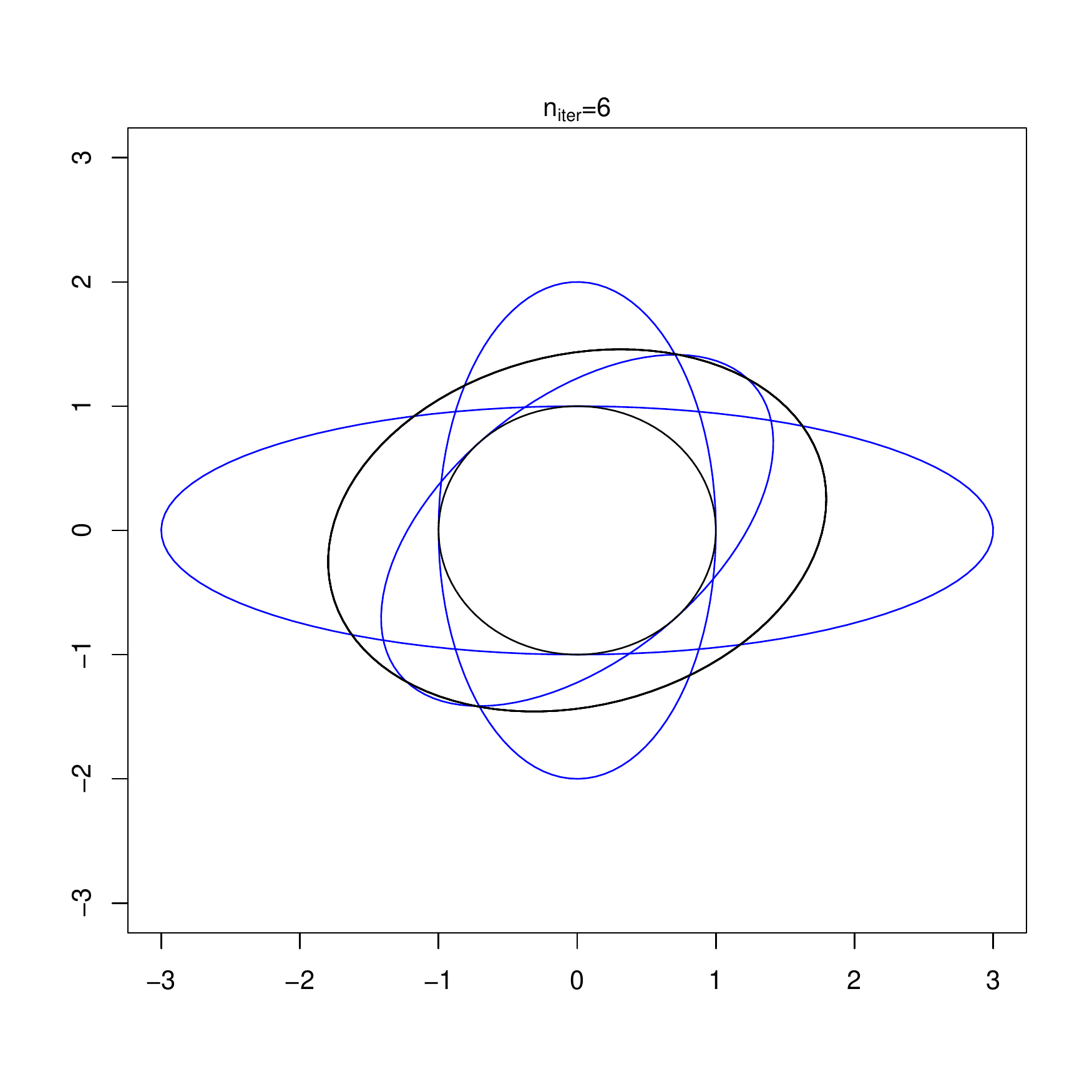}\includegraphics[scale=0.37]{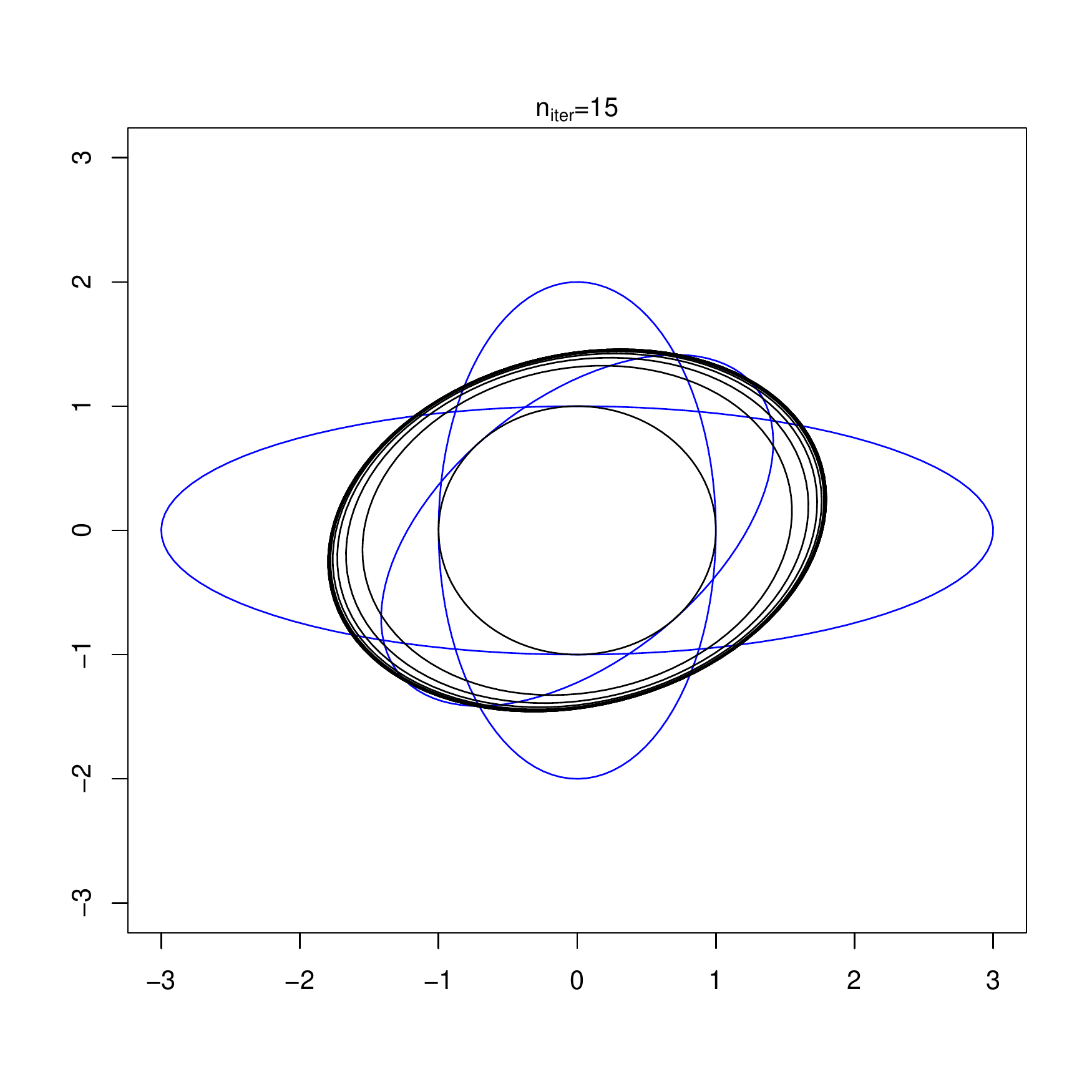}}
\centerline{\includegraphics[scale=0.37]{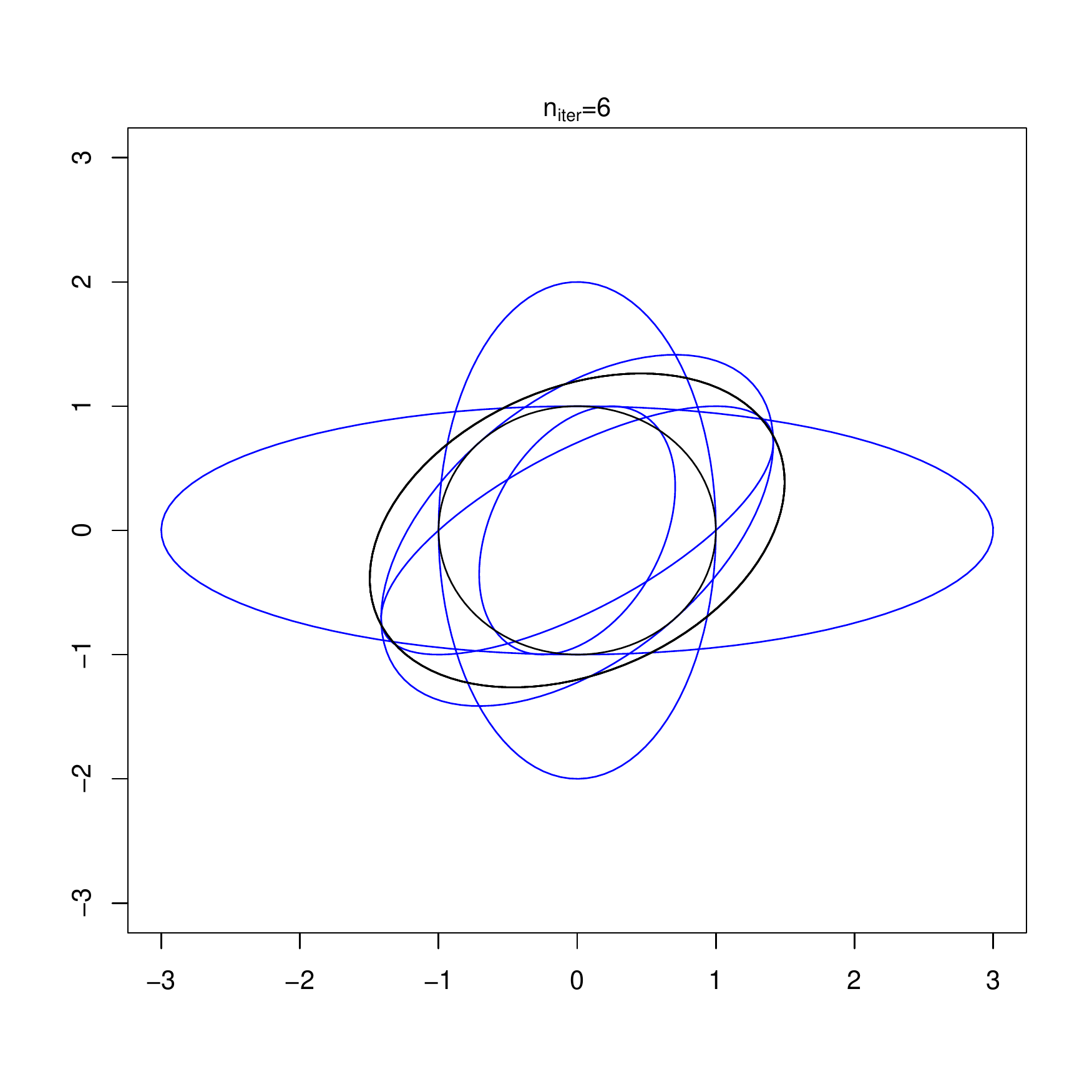}\includegraphics[scale=0.37]{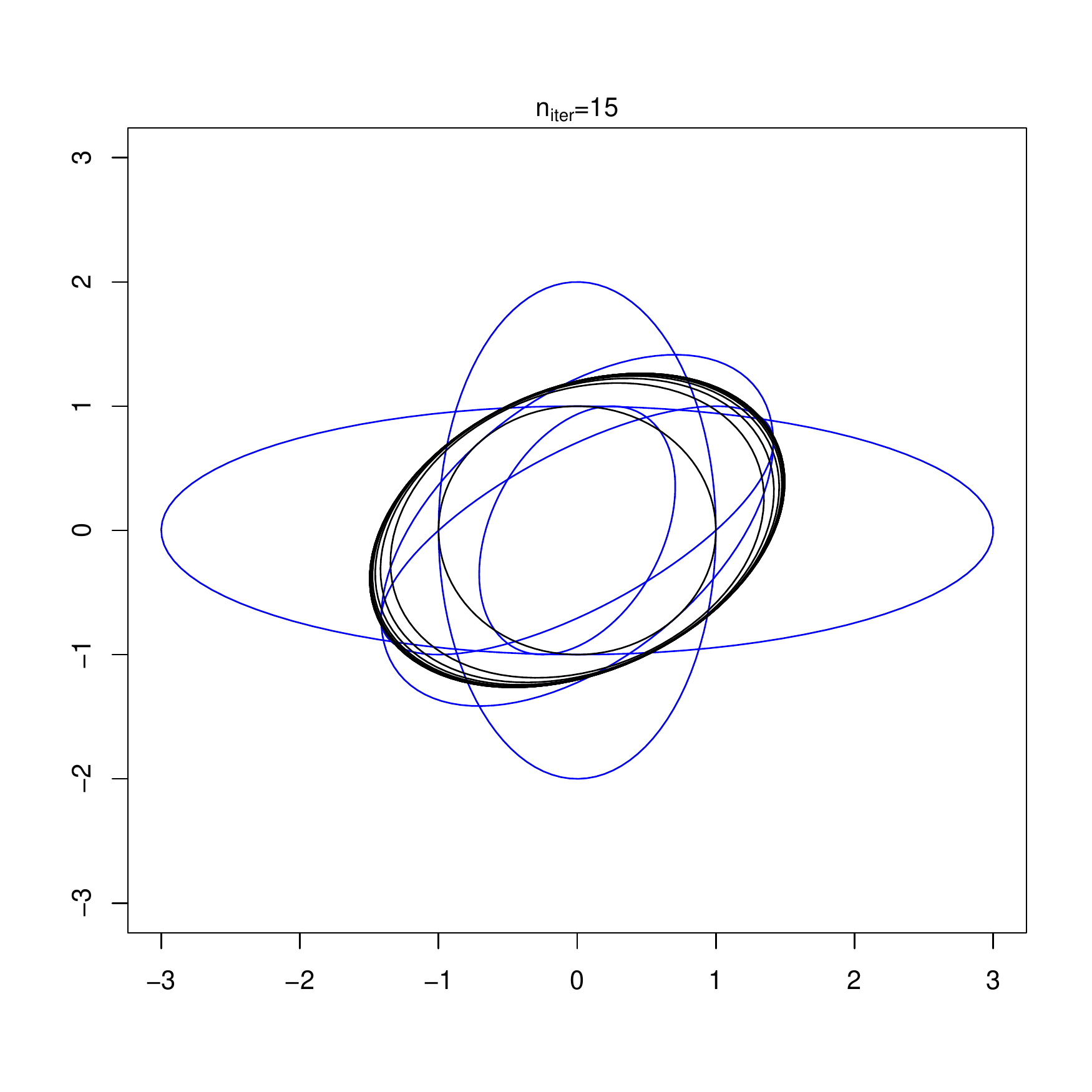}}
\caption{Computation of Wasserstein barycenters, bivariate case. Left: iteration (\ref{iterationnormal}); right: (\ref{iterationRU})}
\end{figure}

Figure \ref{casobivariante} shows results for the case of centered, bivariate
Gaussian distributions. We have considered three different setups. In the top row we have considered
$\Sigma_1=\left[ 
\begin{matrix}
9 & 0\\ 
0 & 1
\end{matrix}
\right]$, $\Sigma_2=\left[ 
\begin{matrix}
1 & 0\\ 
0 & 4
\end{matrix}
\right]$ and $\lambda_1=\lambda_2=\frac 1 2$. In the middle row we consider $\Sigma_1$ and $\Sigma_2$ as before, 
$\Sigma_3=\left[ 
\begin{matrix}
2 & 1\\ 
1 & 2
\end{matrix}
\right]$ and $\lambda_1=\lambda_2=\lambda_3=\frac 1 3$, while the bottom row deals with the same $\Sigma_1$, 
$\Sigma_2$ and $\Sigma_3$ plus 
$\Sigma_4=\left[ 
\begin{matrix}
2 & 1\\ 
1 & 1
\end{matrix}
\right]$ and
$\Sigma_5=\left[ 
\begin{matrix}
\frac 1 2 & \frac 1 4 \\ 
\frac 1 4 & 1
\end{matrix}
\right]$ and weights $\lambda_j=\frac 1 5$, $j=1,\ldots,5$.
The ellipses $\tilde{E}(\Sigma_i)=\{x\in\mathbb{R}^2: \, x^t\Sigma_i^{-1}x= 1\}$ are shown 
in blue, while in black we see the iterants $\tilde{E}(S_n)=\{x\in\mathbb{R}^2: \, x^tS_n^{-1}x= 1\}$.
The number of iterations needed to reach the prespecified tolerance is written on top of each graph.
The graphs on the left column correspond to the proposal in this paper, iteration (\ref{iterationnormal}), while
the right column shows the performance of iteration (\ref{iterationRU}). In all cases we have taken $S_0=Id$.

In the top right graph and in agreement with Remark \ref{onestep} (note that $\Sigma_1$ and $\Sigma_2$ conmute), 
we see that iteration (\ref{iterationnormal}) has found the barycenter after just
one step (the displayed value $\mbox{n}_{\mbox{\scriptsize iter}}=2$ comes from the fact that the iterative procedure computes a further
approximation before checking that the decrease in the target function is below the tolerance and stopping). In contrast, it takes 14 iterations
to (\ref{iterationRU}) to reach convergence. Moving away from this particular case of common principal axis, 
in the middle and bottom row we see that our method converges fast (5 iterations) to the barycenter, again
with somewhat slower convergence for the iteration (\ref{iterationRU}).

For a more complete evaluation of the performance of iteration (\ref{iterationnormal}), as well as comparison
to (\ref{iterationRU}), we have considered a simulation setup with several choices of dimension, $d=2,3,5,10,20,50$
and number of Gaussian distributions, $k=2,3,5,10,20$. For each combination of $d$ and $k$ we have considered
randomly generated, $d\times d$ positive definite matrices $\Sigma_1,\ldots,\Sigma_k$. Each $\Sigma_j$ is generated from a Wishart $W_d(Id,d)$ distribution,
independently of the others. We have used the iteration until convergence (again with tolerance $10^{-10}$) and have recorded the required number
of iterations. The whole procedure has been replicated 1000 times. In Table \ref{tablakd} we report the average number of iterations needed for convergence
for each combination of $d$ and $k$. The left table corresponds to our proposal (\ref{iterationnormal}) and the right table to (\ref{iterationRU}).

We see that, while the number of iterations grows with dimension, in the case of iteration (\ref{iterationnormal}) this number remains moderate, even more so for large values of $k$. We also see that the number of iterations tends to be smaller for larger $k$ (of course, 
the computational cost of each step is higher as both $d$ and $k$ increase). In contrast, iteration (\ref{iterationRU}) is worse affected
by an increase in dimensionality and its performance is uniformly worse than that of (\ref{iterationnormal}).

\begin{table}[ht]
\begin{center}
\begin{tabular}{|c|rrrrr|}
\hline
&\multicolumn{5}{|c|}{$k$}\\
$d$ & 2 & 3 & 5 & 10 & 20
\\
\hline
2 & 6.8 & 7.7 & 7.8 & 7.7 & 7.5\\
3 & 9.3 &10.1 & 10.0 &  9.3 & 8.6\\
5 & 12.2& 13.1& 12.6 & 11.2 & 9.9\\
10 & 17.5& 18.3& 16.4& 13.3 &11.3\\
20 & 22.7& 24.1& 20.4& 15.3 &12.3\\
50 & 31.8& 33.6& 25.7& 17.3 &13.5\\
\hline
\end{tabular}
\quad \quad
\begin{tabular}{|c|rrrrr|}
\hline
&\multicolumn{5}{|c|}{$k$}\\
$d$ & 2 & 3 & 5 & 10 & 20
\\
\hline
2 & 20.6 & 19.1 & 17.6 & 16.4 & 15.7\\
3 & 29.1 & 24.0 & 21.3 & 18.8 & 17.5\\
5 & 41.0 & 31.3 & 25.6 & 21.6 & 19.4\\
10 & 66.9 & 44.2 & 32.0 & 25.2 & 21.9\\
20 & 106.9 & 58.9 & 38.6 & 28.4 & 23.9\\
50 & 185.5 & 82.4 & 47.7 & 32.1 & 26.5\\
\hline
\end{tabular}
\end{center}
\caption{Average iteration numbers for convergence. Left: iteration (\ref{iterationnormal}); right: (\ref{iterationRU})}\label{tablakd}
\end{table}

\begin{figure}[ht]
\centering
\includegraphics[scale=0.6]{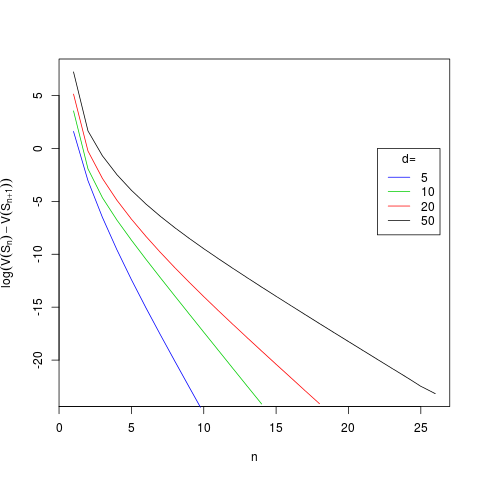}
\caption{log-decrease of target function for iteration (\ref{iterationnormal}), different dimensions $d$}
\label{fastrate}
\end{figure}

Finally, we would like to remark that our numerical experiments show a fast rate of convergence of iteration (\ref{iterationnormal})
to the Wasserstein barycenter. To illustrate this we have included Figure \ref{fastrate}. For this graph we 
have considered dimensions $d=5,10,20,50$. For each of these dimensions we have randomly generated (again from a Wishart distribution) 
five positive definite matrices and have used iteration (\ref{iterationnormal}). We show the decrease of $\log(V(S_n)-V(S_{n+1}))$
with the number of iterations, $n$. We see that in al cases $\log(V(S_n)-V(S_{n+1}))$ shows a linear decrease with $n$. Since
$V(S_n)-V(S_{n+1})$ is an upper bound for $\mathcal{W}_2^2(N(0,S_n),N(0,S_{n+1})$, this indicates an exponential decrease
of $\mathcal{W}_2(N(0,S_n),N(0,S_{n+1})$ and, consequently, exponentially fast convergence of $N(0,S_n)$ towards
the barycenter $N(0, \Sigma_0)$. We believe that this fast convergence of the iterative procedure in this paper deserves further
research that will be reported elsewhere.
\vspace{5mm}

\noindent
{\large \bf Acknowledgements}

\noindent
We want to express our sincere acknowledgements to a referee by his/her constructive and right remarks on our manuscript, leading to a notable improvement of the paper.

\end{document}